\documentclass[3p,onecolumn]{elsarticle}

\usepackage{amssymb,amsmath,mathtools}
\usepackage{graphicx}
\usepackage[dvipsnames]{xcolor}
\usepackage{subfig}
\usepackage{orcidlink}

\usepackage{hyperref}
\hypersetup{                    %
    colorlinks  = true,         %
    linkcolor   = MidnightBlue, %
    citecolor   = MidnightBlue, %
    urlcolor    = MidnightBlue  %
}

\usepackage[american]{babel}
\usepackage{microtype}
\usepackage{bm}
\usepackage{xspace}
\newcommand{\eV}{\text{e\kern-0.15ex V}\xspace} 
\newcommand{\GeV}{\text{G}\eV}
\renewcommand{\v}[1]{\mathbf{#1}}

\begin{document}

\begin{frontmatter}

\title{A Prototype Hybrid Mode Cavity for Heterodyne Axion Detection}

\author[SLAC]{Zenghai Li}
\ead{lizh@slac.stanford.edu}
\author[UCB,LBNL]{\corref{cor}Kevin Zhou\,\orcidlink{0000-0002-9810-3977}}
\ead{kzhou7@berkeley.edu}
\cortext[cor]{Corresponding author}
\author[SLAC]{Marco Oriunno}
\ead{oriunno@slac.stanford.edu}
\author[FNAL,SQMS]{Asher Berlin\,\orcidlink{0000-0002-1156-1482}}
\ead{aberlin@fnal.gov}
\author[CERN]{Sergio Calatroni\,\orcidlink{0000-0002-2769-8029}}
\ead{sergio.calatroni@cern.ch}
\author[IPhT,ENS]{Raffaele Tito D'Agnolo}
\author[Unige]{Sebastian A.~R.~Ellis\,\orcidlink{0000-0003-3611-2437}}
\ead{sebastian.ellis@unige.ch}
\author[SLAC]{Philip Schuster}
\ead{schuster@slac.stanford.edu}
\author[SLAC,ASU]{Sami G.~Tantawi}
\ead{stantawi@asu.edu}
\author[SLAC]{Natalia Toro}
\ead{ntoro@slac.stanford.edu}

\affiliation[SLAC]{organization={SLAC National Accelerator Laboratory},
            addressline={2575 Sand Hill Road}, 
            city={Menlo Park},
            postcode={94025}, 
            state={CA},
            country={USA}}
\affiliation[UCB]{organization={Berkeley Center for Theoretical Physics},
            addressline={University of California},
            city={Berkeley},
            postcode={94720},
            state={CA},
            country={USA}}
\affiliation[LBNL]{organization={Theoretical Physics Group},
            addressline={Lawrence Berkeley National Laboratory},
            city={Berkeley},
            postcode={94720},
            state={CA},
            country={USA}}
\affiliation[FNAL]{organization={Theoretical Physics Division, Fermi National Accelerator Laboratory},
            city={Batavia},
            postcode={60510}, 
            state={IL},
            country={USA}}
\affiliation[SQMS]{organization={Superconducting Quantum Materials and Systems Center (SQMS), Fermi National Accelerator Laboratory},
            city={Batavia},
            postcode={60510}, 
            state={IL},
            country={USA}}
\affiliation[CERN]{organization={CERN},
            addressline={1211 Geneva 23},
            country={Switzerland}}
\affiliation[IPhT]{Institut de Physique Theorique, Universite Paris Saclay, CNRS CEA, F-91191 Gif-sur-Yvette, France}
\affiliation[ENS]{Laboratoire de Physique de l’Ecole Normale Superieure, ENS, Universite PSL, CNRS, Sorbonne
Universite, Universite Paris Cite, F-75005 Paris, France}
\affiliation[Unige]{organization={Departement de Physique Theorique, Universite de Geneve},
            addressline={24 quai Ernest Ansermet}, 
            city={Geneve},
            postcode={1211}, 
            country={Switzerland}}
\affiliation[ASU]{organization={Arizona State University},
            addressline={797 E. Tyler Street}, 
            city={Tempe},
            postcode={85281}, 
            state={AZ},
            country={USA}}

\begin{abstract}
In the heterodyne approach to axion detection, axion dark matter induces transitions between two modes of a microwave cavity, resulting in a parametrically enhanced signal power. We describe the fabrication and characterization of a prototype normal conducting cavity specifically optimized for heterodyne detection. Corrugations on the cavity walls support linearly polarized hybrid modes which maximize the signal power while strongly suppressing noise. We demonstrate tuning mechanisms which allow one mode frequency to be scanned across a 4 MHz range, while suppressing cross-coupling noise by at least 80 dB. A future superconducting cavity with identical geometry to our prototype would have the potential to probe orders of magnitude beyond astrophysical bounds.
\end{abstract}


\end{frontmatter}

\section{Introduction}
\label{sec:intro}

Axions have emerged as a leading dark matter candidate. They are motivated both by their minimality and simplicity, and by their generic appearance in theories beyond the Standard Model. The search for axion dark matter is promising because axions have a small number of canonical interactions, which can be probed with great sensitivity in new precision experiments. 

Most ideas for axion detection use the fact that in the presence of a magnetic field $\v{B}$, the axion sources an effective electric current $\v{J}_{\mathrm{eff}} \simeq g_{a\gamma\gamma} \dot{a} \v{B}$ which in turn sources detectable electromagnetic fields. Though many new detection concepts have been proposed in the past decade~\cite{Irastorza:2018dyq,Sikivie:2020zpn,Berlin:2024pzi}, to date only cavity haloscopes have achieved significant sensitivity beyond astrophysical bounds. In this class of experiments, originally conceived by Sikivie in the 1980s~\cite{Sikivie:1983ip,Sikivie:1985yu}, a large static magnetic field is applied, and the oscillating effective current resonantly excites a mode of a microwave cavity. This approach is well-suited to probing axions with GHz frequencies, corresponding to a mass
\begin{equation}
m_a \simeq 4 \times 10^{-6} \, \frac{\eV}{c^2} \, \left( \frac{f}{\mathrm{GHz}} \right)
\end{equation}
and a symmetry breaking scale of $\sim 10^{12} \, \GeV$~\cite{GrillidiCortona:2015jxo}. However, significantly lower axion masses of $m_a \lesssim 10^{-8} \, \eV/c^2$ are motivated by string theory and grand unification~\cite{Banks:2003sx,Svrcek:2006yi,Arvanitaki:2009fg,Stott:2017hvl,Broeckel:2021dpz,Demirtas:2021gsq,Gendler:2023kjt,Benabou:2025kgx}. In this regime, a traditional cavity haloscope would become impractically large. 

This problem can be circumvented using the heterodyne approach to axion detection, in which a microwave cavity is excited in a ``loaded'' mode of angular frequency $\omega_0$. This causes the axion's effective current to oscillate at angular frequency $\omega_0 \pm \omega_a$, where $\omega_a = m_a c^2 / \hbar$ is the axion's angular frequency. The effective current can resonantly drive a ``signal'' mode of angular frequency $\omega_1$ provided that $\omega_a = |\omega_0 - \omega_1|$. A range of axion masses can be probed by tuning the frequency difference between these modes. The axion signal power is proportional to the energy stored in the loaded mode, which can be maintained at a high value in a superconducting cavity.

This concept could be used to search for axions with sub-Hz to GHz frequencies. However, in addition to the high signal power, there is also noise power due to ``leakage'' from the loaded mode to the signal mode, which can result from the cross-couplings $\chi_{\text{d}, \text{r}}$ (with $\chi_{\text{d}}$ and $\chi_{\text{r}}$ associated with the driving and readout waveguides, respectively) or from mode mixing induced by mechanical vibrations, characterized by mechanical form factors $\eta_p$. These noise sources fall rapidly as the axion frequency increases above $\sim$ kHz. In addition, the tuning range of a typical cavity would be of order $\sim$ MHz. Thus, the heterodyne approach is particularly powerful for $\sim$ kHz to $\sim$ MHz axion frequencies, which are the focus of this work.

To optimize a heterodyne experiment, one must design a cavity that maximizes the energy that can be stored in the loaded mode, and the dimensionless overlap $C_{\text{sig}}$ of the effective current with the signal mode. The mode frequencies should be tunable over a wide range, while maintaining low values for $\chi_{\text{d}, \text{r}}$ and $\eta_p$ to suppress leakage noise. Many of these considerations are unique to the heterodyne approach, and require a dedicated study. 

\begin{figure}
\centering
\subfloat[\label{fig:cavity_outside}]{\includegraphics[width=0.553\linewidth]{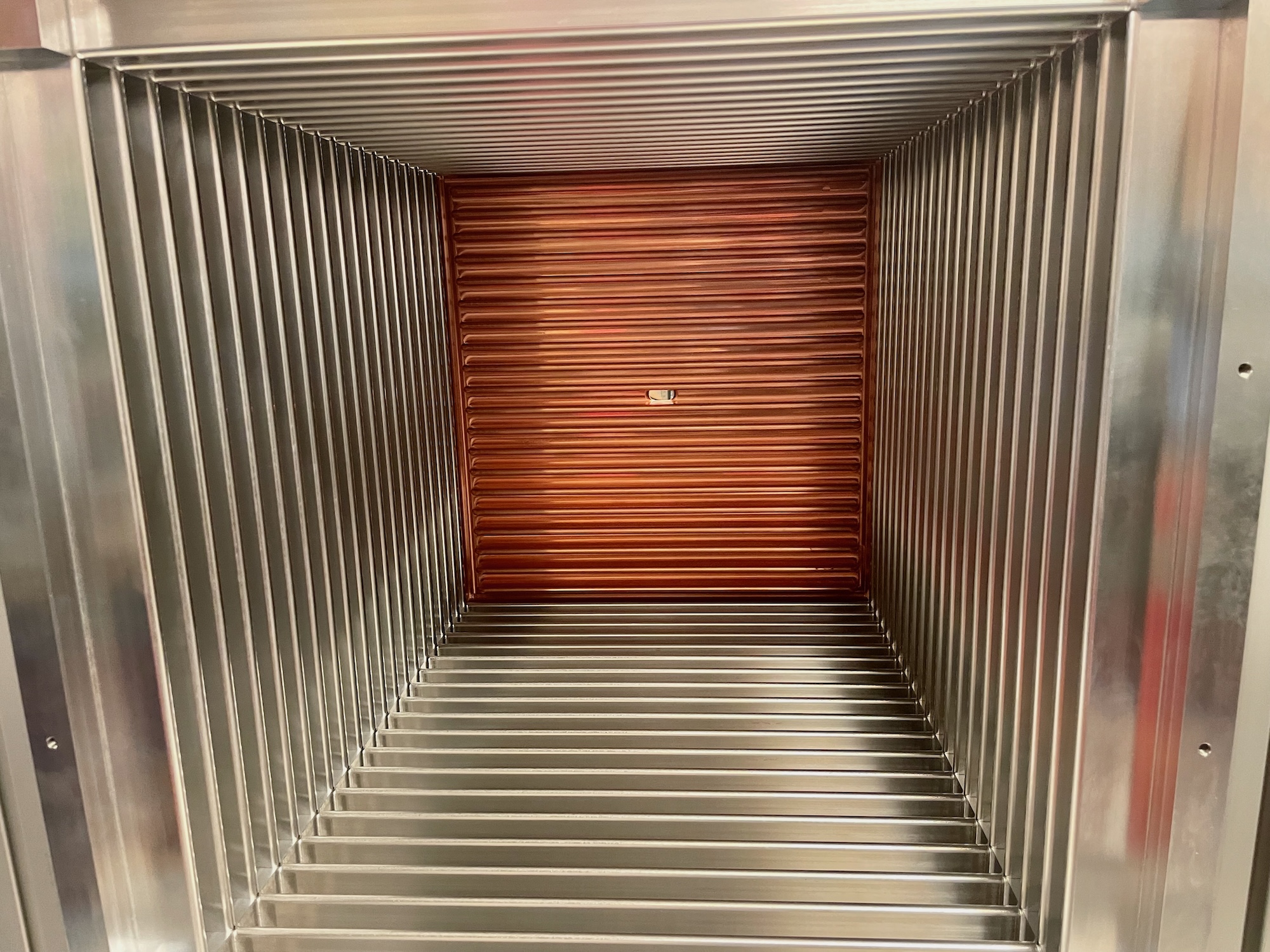}} \quad
\subfloat[\label{fig:cavity_inside}]{\includegraphics[width=0.4\linewidth]{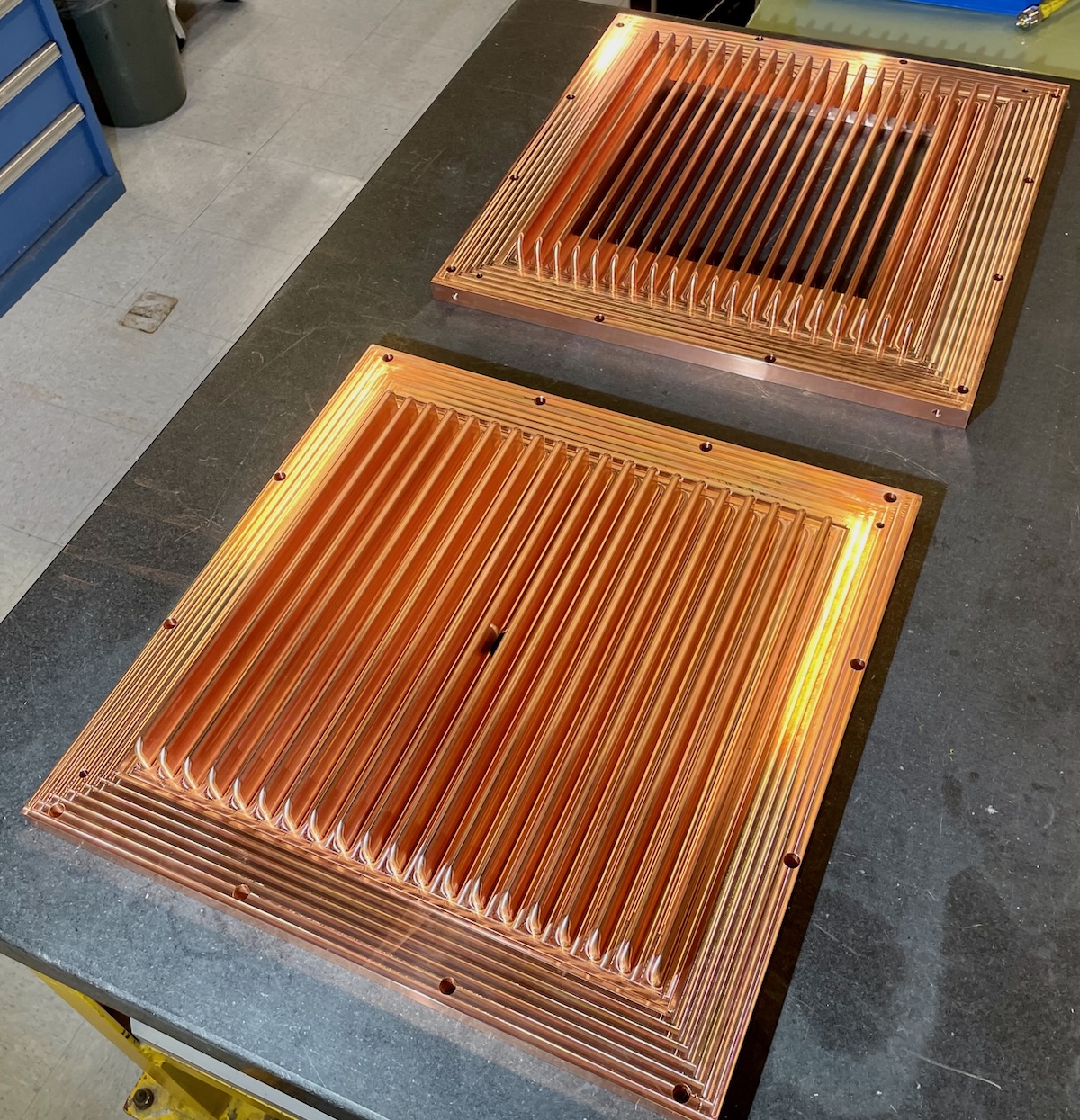}}
\vfill
\caption{Geometry of the prototype cavity. (a) Interior of the cavity, showing corrugations on the aluminum side walls and the ``fixed'' copper endplate. (b) The corrugated copper endplates. The ``tunable'' endplate is open at the back, so that a tuning plate (shown in Fig.~\ref{fig:tuner_fab}) can be deformed behind it, adjusting the mode frequency difference.}
\label{fig:cavity}
\end{figure}

In this work, we describe the design, fabrication, and characterization of a prototype cavity at SLAC whose geometry is optimized for heterodyne axion detection. The cavity, shown in Fig.~\ref{fig:cavity}, is fabricated from six rectangular plates. Due to corrugations on the walls, the cavity carries a pair of nearly degenerate, approximately linearly polarized $\mathrm{HE}_{11}$ hybrid modes. Two corrugated endplates shift the modes relative to each other, leading to an almost ideal signal overlap factor $C_{\text{sig}}$ and suppressed values of $\eta_p$. Ports for the driving and readout waveguides are placed inside opposite endplates, where the signal and loaded modes are excluded respectively, automatically leading to a strong suppression of $\chi_{\text{d}, \text{r}}$. One endplate can be deformed to tune the mode frequency difference, and rotated to further decrease $\chi_{\text{d}, \text{r}}$.

Since our proof-of-principle cavity is not superconducting, it has limited sensitivity to axion dark matter, and we do not plan to use it for a dark matter search. However, our simulations and measurements demonstrate that our novel cavity design can simultaneously maximize signal and suppress noise, while maintaining a wide tuning range. In addition, our design can be fabricated out of flat plates, which will simplify the transition to a superconducting niobium cavity in future work. 

The paper is organized as follows. Sec.~\ref{sec:heterodyne} reviews the heterodyne approach to axion detection, and defines the form factors that determine its signal and noise. (This section uses natural units, $\hbar = c = 1$.) Sec.~\ref{sec:cavity_design} reviews the theory of hybrid modes, describes the cavity's design, and shows numeric calculations of the mode profiles, $\chi_{\text{d}, \text{r}}$, and $\eta_p$. Sec.~\ref{sec:mech_fab} describes the mechanical fabrication of the cavity and its tuning mechanism. Sec.~\ref{sec:cavity_test} presents results from bench measurements of the finished cavity, which corroborate the results of simulations. In particular, we demonstrate suppression of the cross-coupling down to at least $\chi_{\text{d}, \text{r}} \sim 10^{-4}$, and a tuning range of $4 \, \mathrm{MHz}$. In Sec.~\ref{sec:conclusion}, we review related work and discuss potential next steps.

\section{Heterodyne Axion Detection}
\label{sec:heterodyne}

The heterodyne approach to axion detection was first considered in 2010 by Sikivie~\cite{Sikivie:2010fa}, who focused on the case of $\sim \mathrm{GHz}$ axions, in the regime $\omega_0 \ll \omega_1 \approx m_a$. In this case, it was found that heterodyne detection did not provide advantages over the traditional cavity haloscope. A decade later, it was realized that for lighter axions, $m_a \ll \omega_0 \approx \omega_1$, the heterodyne approach yields a parametrically enhanced signal power compared to alternatives using a static background field. Several works analyzed the potential sensitivity of this approach~\cite{Berlin:2019ahk,Lasenby:2019prg,Berlin:2020vrk}, exploring cavity designs and noise sources.\footnote{Earlier, Refs.~\cite{Goryachev:2018vjt,PhysRevLett.126.081803} (with corrections in Refs.~\cite{THOMSON2021100787,PhysRevLett.127.019901}) proposed driving two modes of a cavity and measuring frequency shifts induced by axion dark matter. We will call this the ``frequency technique'', in contrast to the ``power technique'' which is the focus of this work.} Here we will briefly review the factors that determine the signal and noise, to establish notation for the following sections. 

\subsection{Signal Power}

To derive the signal power in a heterodyne axion experiment, let the relevant modes have electric and magnetic field profiles $\tilde{\v{E}}_i(\v{r})$ and $\tilde{\v{B}}_i(\v{r})$, where $i = 0$ for the loaded mode and $i = 1$ for the signal mode. These profiles are normalized so that the integrals of $|\tilde{\v{E}}_i|^2$ and $|\tilde{\v{B}}_i|^2$ over the cavity volume $V$ are all equal to one. Exciting the loaded mode produces an oscillating background magnetic field $\v{B}_0 = b_0(t) \tilde{\v{B}}_0(\v{r})$ in the cavity. The electric field of the signal mode is $e_1(t) \tilde{\v{E}}_1(\v{r})$, and Maxwell's equations imply that its amplitude obeys
\begin{equation}
\omega_1^2 e_1 + \frac{\omega_1}{Q_1} \, \dot{e}_1 + \ddot{e}_1 = -\! \int_V \tilde{\v{E}}_1^* \cdot \dot{\v{J}}_{\mathrm{eff}}
\end{equation}
where $\v{J}_{\mathrm{eff}} \simeq g_{a\gamma\gamma} \dot{a} \v{B}_0$. We define the signal mode lifetime $t_1 = Q_1/\omega_1$, and for simplicity assume it is much shorter than the axion coherence time. Then when the axion drives the signal mode on resonance, $m_a \simeq |\omega_0 - \omega_1|$, the signal amplitude is 
\begin{equation}
e_1 \sim t_1 \, (g_{a\gamma\gamma} \, \dot{a} \, b_0) \int_V \tilde{\v{E}}_1^* \cdot \tilde{\v{B}}_0 
\end{equation}
where $\sim$ indicates equality up to a dimensionless constant. In the steady state, the power driven into the signal mode is 
\begin{equation} \label{eq:signal_power}
P_{\text{sig}} \sim \frac{e_1^2}{t_1} \sim g_{a\gamma\gamma}^2 \rho_{_\mathrm{DM}} \, B_0^2 V \, C_{\text{sig}}^2 \, t_1
\end{equation}
where we used the fact that the dark matter density is $\rho_{_\mathrm{DM}} \sim \dot{a}^2$, we let $b_0^2 = B_0^2 V$ to match the notation of Ref.~\cite{Berlin:2019ahk}, and defined the signal form factor 
\begin{equation} \label{eq:signal_form}
C_{\text{sig}} = \bigg| \int_V \tilde{\v{E}}_1^* \cdot \tilde{\v{B}}_0 \bigg| \leq 1.
\end{equation}
Note that Eq.~\eqref{eq:signal_power} holds when the axion coherence time is longer than $t_1$, i.e.~when the axion is narrower in frequency than the signal mode. For a more detailed derivation of the general result, see Ref.~\cite{Berlin:2019ahk}.

We can read off several distinctive properties of the heterodyne approach from Eq.~\eqref{eq:signal_power}. First, for fixed axion density, the signal power is independent of the axion mass. This provides a strong parametric advantage over an approach using a static magnetic field. In that case, the electric field profile near the magnetic field scales as $\tilde{\v{E}}_1 \sim m_a L \,\tilde{\v{B}}_1$ in the magnetoquasistatic limit, leading to $C_{\text{sig}} \sim m_a L$ and therefore a signal power suppressed by $P_{\text{sig}} \propto (m_a L)^2$ whenever $m_a L \ll 1$. This suppression holds very generally, applying to reentrant cavities~\cite{Chakrabarty:2023rha}, dielectric loaded cavities~\cite{Chakrabarty:2023rha,Bai:2023ajm}, and LC circuits~\cite{DMRadio:2022pkf}. By avoiding this penalty, the heterodyne approach starts out with an enhancement of $\sim 10^6$ in signal power at $\mathrm{MHz}$ frequencies.

Second, an effective heterodyne experiment must use a superconducting cavity. This is because maintaining a high loaded mode amplitude $b_0$ requires a high quality factor $Q_0$ for the loaded mode, and allowing the signal to build up for a sufficient time $t_1$ requires a high quality factor $Q_1$ for the signal mode. However, to preserve superconductivity, a heterodyne experiment must operate at a lower background magnetic field than a static field experiment. In addition, it must operate at a higher temperature, $T \sim 2 \, \mathrm{K}$, to handle the energy dissipation in the walls. However, these disadvantages are compensated by the advantages of higher quality factor and enhanced signal power. 


\subsection{Noise Sources}

At sufficiently high axion frequencies, the dominant noise sources in a heterodyne axion experiment are thermal noise, set by the temperature $T \simeq 2 \, \mathrm{K}$ of the liquid helium bath, and amplifier noise. (For concreteness, the thermal occupancy of a GHz frequency mode at this temperature is $T / (2 \pi \, \mathrm{GHz}) \simeq 40$.)

There are also additional noise sources proportional to the power $P_{\text{in}}$ input to the cavity, which become increasingly important at lower axion frequencies. This ``leakage'' occurs either because the waveguides used to drive and readout the cavity couple partially to the wrong mode (``cross-coupling''), or because vibrations directly transfer energy from the loaded mode to the signal mode (``mode mixing''). We consider these effects in the next two subsections. In both cases, the noise power is suppressed both by a geometric form factor, and by a factor depending on the mode separation $m_a = |\omega_0 - \omega_1|$, which falls off at higher $m_a$.  

\subsubsection{Cross-Coupling}
\label{subsec:cross_coupling}

In general, the noise power due to cross-coupling is proportional to the input power $P_{\text{in}}$, a cross-coupling parameter $\chi^2$, and to a spectral density which describes how power is converted across the angular frequency separation $m_a$. For example, oscillator phase noise contributes a noise power $\sim P_{\text{in}} (\omega_0 / Q_0) \, \chi^2 \, S_\varphi(m_a)$, where $S_\varphi$ is the phase noise spectral density, with units of $1/\mathrm{Hz}$. There are also contributions due to oscillator amplitude noise (though we expect it to be subdominant to phase noise at low $m_a$), and from mechanical vibrations which shift the cavity mode frequencies; see Ref.~\cite{Berlin:2019ahk} for further discussion. Since these quantities depend on, e.g.~the quality of the high power driving oscillator and vibrations in the cavity's environment, they cannot be directly characterized in our prototype. Thus, we will focus on the cross-coupling, which depends only on the cavity geometry. 

To quantify it, let $Q_i$ be the loaded quality factor for the loaded ($i = 0$) and signal ($i = 1)$ mode. The loss rate in each mode is proportional to $1/Q_i$, and can be decomposed into an intrinsic loss and a loss due to the couplings to the driving and readout waveguides (quantified by external quality factors $Q_{i\text{d}}$ and $Q_{i\text{r}}$), so that 
\begin{equation}
\frac{1}{Q_i} = \frac{1}{Q_i^{\text{int}}} + \frac{1}{Q_{i\text{d}}} + \frac{1}{Q_{i\text{r}}}.
\end{equation}
Here, $Q_i^{\text{int}}$ is the intrinsic quality factor, i.e.~the quality factor that mode $i$ would have in the absence of the waveguides. Ideally, the driving waveguide couples only to the loaded mode, and the readout waveguide couples only to the signal mode. The strength of this coupling is parametrized by the RF coupling strengths $\beta_0 = Q_0^{\text{int}} / Q_{0\text{d}}$ and $\beta_1 = Q_1^{\text{int}} / Q_{1\text{r}}$. There are also unwanted cross-couplings, which we parametrize by
\begin{equation}
\chi_{\text{d}}^2 = \frac{1/Q_{1\text{d}}}{1/Q_{0\text{d}}}, \qquad \chi_\text{r}^2 = \frac{1/Q_{0\text{r}}}{1/Q_{1\text{r}}}.
\end{equation}
One would generically expect the typical cross-coupling $\chi \sim \chi_\text{d} \sim \chi_\text{r}$ to be roughly set by the fractional precision with which one can mechanically align a waveguide antenna. Stronger suppressions can be achieved using special cavity geometries. For instance, the MAGO collaboration used a two-sphere cavity coupled to a ``magic tee'' with an adjustable phase shift to achieve $\chi \sim 10^{-7}$, corresponding to $140 \, \text{dB}$ of noise reduction~\cite{Ballantini:2005am}. Since $S_f(m_a)$ decreases steeply as the axion mass increases, Ref.~\cite{Berlin:2019ahk} estimated\footnote{In that work, the quantities $\chi_{\text{d}}$ and $\chi_{\text{r}}$ were denoted as $\epsilon_{1\text{d}}$ and $\epsilon_{0\text{r}}$, respectively.} it becomes subdominant to thermal noise for axion frequencies above $\sim \text{kHz}$ for $\chi \sim 10^{-7}$, and above $\sim \text{MHz}$ for $\chi \sim 10^{-5}$.

Here, we will infer the cross-couplings using microwave scattering parameters $S_{ij}(\omega)$. (For a pedagogical introduction to $S$-parameters, see Ref.~\cite{pozar}.) We critically couple both waveguides, $Q_{0\text{d}} \simeq Q_0^{\text{int}}$ and $Q_{1\text{r}} \simeq Q_1^{\text{int}}$, corresponding to $\beta_0 = \beta_1 \simeq 1$, so that when the cavity is driven with power $P_{\text{in}}$ through the driving waveguide, on resonance with the loaded mode, no power is reflected, $S_{\text{dd}}(\omega_0) \approx 0$. For $\chi_{\text{r}} \ll 1$, a power $\chi_{\text{r}}^2 \, P_{\text{in}}$ exits through the readout waveguide, so we can identify
\begin{equation} \label{eq:cross_coupling_def}
\chi_{\text{r}}^2 \simeq |S_{\text{dr}}(\omega_0)|^2, \qquad \chi_{\text{d}}^2 \simeq |S_{\text{rd}}(\omega_1)|^2,
\end{equation}
where the second equality follows from similar reasoning. Note that since the cavity is a reciprocal network, we have $S_{\text{rd}}(\omega) = S_{\text{dr}}(\omega)$. Thus, at each input frequency there is only one physical cross-coupling. In a heterodyne experiment, cross-coupling noise is predominantly due to the signal mode being driven through $\chi_\text{d}$, and the associated noise power is proportional to $\chi_\text{d}^2$.

\subsubsection{Mechanical Mode Mixing}

When the loaded mode is excited, the cavity walls carry surface currents oscillating with angular frequency $\omega_0$. A vibration of the walls in a mechanical mode $p$ with angular frequency $\omega_p$ thus leads to a current inside the cavity with angular frequency $\omega_0 \pm \omega_p$, which can resonantly excite the signal mode when $\omega_p \simeq |\omega_0 - \omega_1|$. The efficiency of this process is determined by the dimensionless form factor~\cite{Berlin:2023grv}
\begin{equation} \label{eq:eta_mix}
\eta_p = V^{1/3} \, \bigg| \int_{S} d\v{S} \cdot \bm{\xi}_p \, (\tilde{\v{E}}_0^* \cdot \tilde{\v{E}}_1 - \tilde{\v{B}}_0^* \cdot \tilde{\v{B}}_1) \bigg| \, ,
\end{equation}
where $S$ is the cavity's inner surface, the approximation $\omega_0 \simeq \omega_1$ was made, and $\bm{\xi}_p$ is the dimensionless spatial profile of the mechanical mode, normalized so that the average value of $\xi_p^2$ within the shell of the cavity is $1$. At the $\sim \mathrm{kHz}$ to $\sim \mathrm{MHz}$ frequencies relevant for heterodyne detection, the energy in mechanical modes is largely determined by the spectrum of external vibrations. Explicitly, a mechanical mode with $\omega_p \simeq m_a$ carrying a fraction $U_p$ of the total vibrational energy contributes a mechanical mode mixing noise power proportional to $P_{\text{in}} \, \eta_p^2 \, U_p$. 

For a generic cavity geometry, one would expect that $\eta_p$ is order-one for low-lying mechanical modes, and suppressed for generic higher frequency mechanical modes, as their profiles $\bm{\xi}_p$ would rapidly oscillate across the cavity surface. For special cavity geometries, $\eta_p$ can be further suppressed. For instance, in Ref.~\cite{Berlin:2019ahk}, it was noted that for certain choices of cylindrical cavity modes, the dot products in~\eqref{eq:eta_mix} always vanish. However, Ref.~\cite{Lasenby:2019prg} pointed out that in this case, $\eta_p$ is proportional to the fractional deviation of the cavity from this ideal shape, e.g.~by an ellipticity. 

It is likely that noise from mechanical mode mixing will greatly dominate over thermal noise at $\text{kHz}$ frequencies, but it falls off rapidly at higher frequencies. This is both because the form factor $\eta_p$ of higher mechanical modes will decrease, and because the mechanical power $P_p$ will decrease due to the falling spectrum of external vibrations. Thus, the overall importance of this noise source depends on how quickly these quantities fall off, and other factors such as the rigidity of a superconducting cavity, and the amount of vibrations sourced by its helium cryostat. 

We cannot address those issues with our prototype, as it is made from non-superconducting materials with different mechanical properties, and sits in air at room temperature. However, we will show that for simple choices of $\bm{\xi}_p$, the form factor $\eta_p$ is highly suppressed due to the properties of the hybrid modes. 

\section{Cavity Design}
\label{sec:cavity_design}

In this section, we discuss the design of our prototype cavity. Sec.~\ref{subsec:concept} discusses how the original conceptual design uses the $\text{HE}_{11}$ hybrid mode to enhance the signal. Sec.~\ref{subsec:optimized} describes the cavity's final optimized design, and Sec.~\ref{subsec:tuners} describes its coupling and frequency tuning. Finally, we numerically compute the cross-coupling $\chi$ in Sec.~\ref{subsec:cross_couple}, and form factors for mechanical mode mixing in Sec.~\ref{subsec:form_factors}. 

\subsection{Hybrid Mode Concept}
\label{subsec:concept}

\begin{figure}[t!]
\centering
\subfloat[\label{fig:cylinder_cross}]{\includegraphics[width=0.6\linewidth]{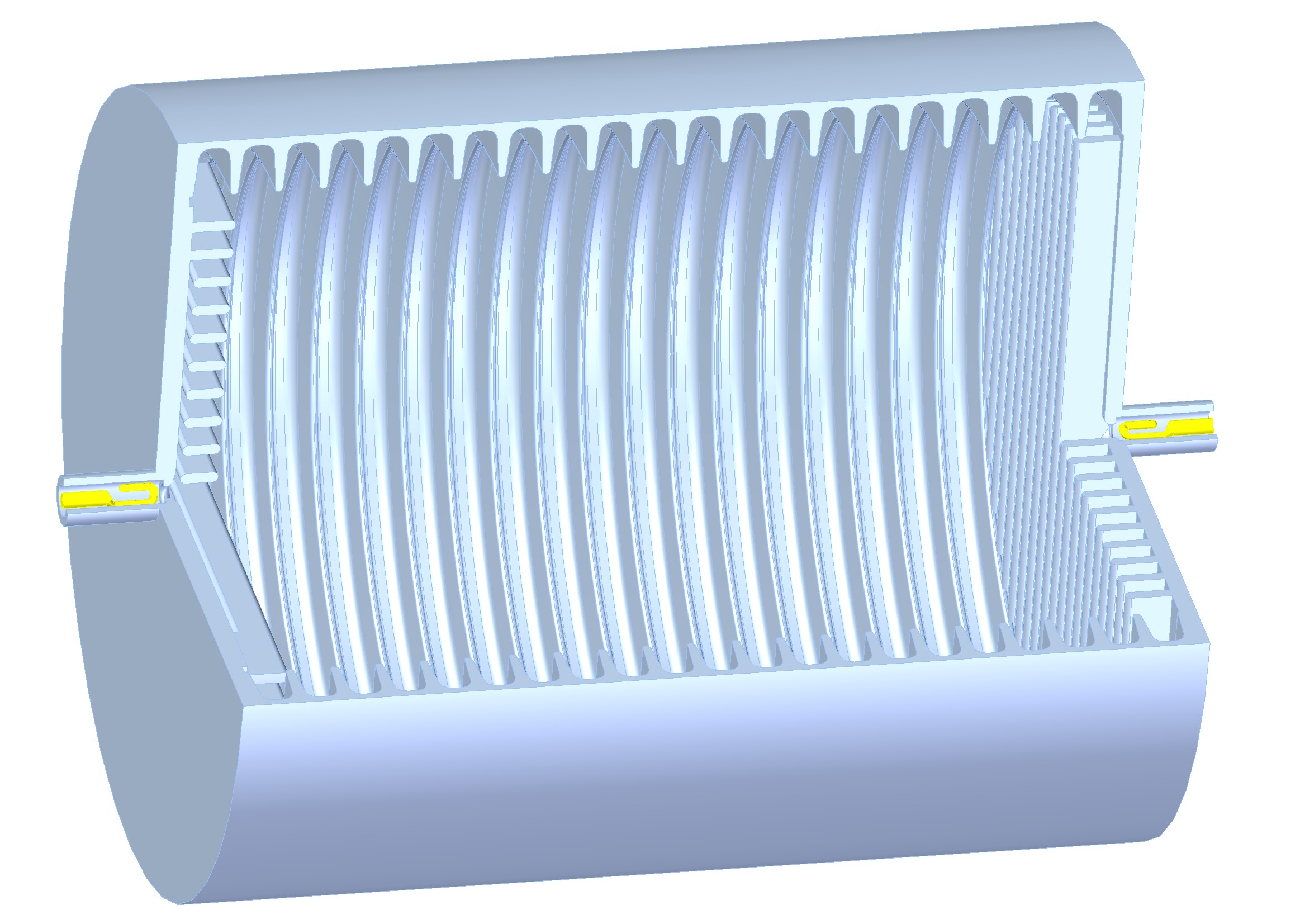}}
\subfloat[\label{fig:eb_cylinder}]{\includegraphics[width=0.38\linewidth]{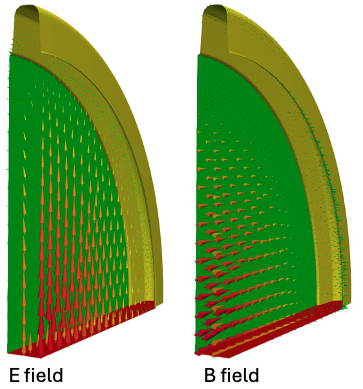}} \newline
\subfloat[\label{fig:modes_2_cylinder}]{\includegraphics[width=0.87\linewidth]{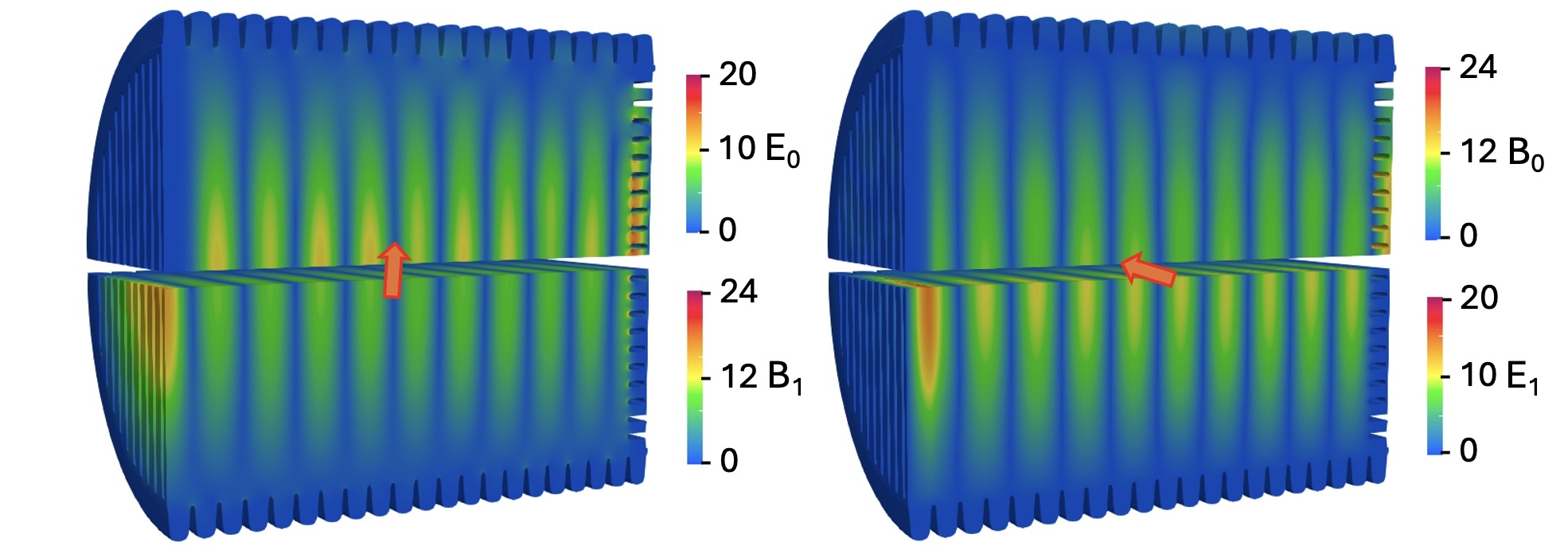}}
\vfill
\caption{Original conceptual design of the hybrid mode cavity. (a) The cavity has corrugated side walls which support highly overmoded $\text{HE}_{11}$ modes, and endplate fins shift the modes by $\lambda/4$ relative to each other. (b) Electric and magnetic field profiles of an $\text{HE}_{11}$ mode in a $\lambda/4$ period. The fields are linearly polarized, orthogonal, and suppressed at the cavity side walls. (c) Full mode profiles of both $\text{HE}_{11}$ modes, in units of $\mathrm{m}^{-3/2}$. Here, $\tilde{\v{E}}_0$ and $\tilde{\v{B}}_1$ are both approximately vertical, while $\tilde{\v{E}}_1$ and $\tilde{\v{B}}_0$ are both approximately horizontal. We show the vertical components of $\tilde{\v{E}}_0$ and $\tilde{\v{B}}_1$ at left, and the horizontal components of $\tilde{\v{E}}_1$ and $\tilde{\v{B}}_0$ at right. They overlap almost perfectly within the cavity, giving a high signal form factor.}
\label{fig:cylinder}
\end{figure}

The original cavity design concept, discussed in Ref.~\cite{Berlin:2019ahk} and shown in Fig.~\ref{fig:cylinder_cross}, involved the two linearly polarized ``hybrid'' $\text{HE}_{11}$ modes in a overmoded corrugated cylindrical cavity. The theory of hybrid modes is discussed in Ref.~\cite{clarricoats1984corrugated}, with modern applications discussed in Refs.~\cite{5597962,Zhang:xe5035}. Here we give a brief introduction. 

In an ordinary cylindrical cavity of radius $R$ and length $L$, the resonant modes are $\text{TE}_{mnp}$ and $\text{TM}_{mnp}$, where the integer $m$ describes the azimuthal variation, $n \geq 1$ describes the radial variation, and $p \geq 0$ describes the variation along the cylinder axis. The index $p$, which we will leave implicit, is related to the wavelength $\lambda$ along the axis by $p \lambda = 2 L$. To achieve a maximal signal form factor $C_{\text{sig}}$, we seek to align the transverse fields of two modes, while minimizing longitudinal fields. The ratio of the magnitudes of the longitudinal and transverse fields are of order $\lambda /R$, so we consider modes in the overmoded limit $\lambda \ll R$.

The transverse electric and magnetic fields of each of the $\text{TE}_{11}$ modes (from taking $m = \pm 1$) roughly point in a single direction, suggesting that they can be arranged to have a high signal form factor. However, one can do even better by corrugating the cavity side walls, so that they effectively present a boundary condition with nontrivial impedance. In this case, the resonant modes are the hybrid modes $\text{HE}_{mn}$ and $\text{EH}_{mn}$, which are constructed by superposing $\text{TE}_{mn}$ and $\text{TM}_{mn}$ profiles. In the ``balanced hybrid'' limit, corresponding to corrugations with depth $\lambda/4$, these profiles are combined with equal or opposite amplitude, so that the transverse electric and magnetic fields of each $\text{HE}_{11}$ mode are perfectly linearly polarized, as shown in Fig.~\ref{fig:eb_cylinder}. The transverse electric and magnetic fields have identical transverse spatial distributions, but point in orthogonal directions, and are offset by $\lambda/4$ along the cavity axis. 

If we directly took the two $\text{HE}_{11}$ modes to be the loaded and signal mode, then $\tilde{\v{E}}_1^*$ would be parallel to $\tilde{\v{B}}_0$ but offset by $\lambda/4$, leading to a cancellation in the signal form factor in Eq.~\eqref{eq:signal_form}. To resolve this problem, fins of depth $\lambda/4$ can be added to the endplates. As shown in Fig.~\ref{fig:cylinder_cross}, each set of fins reflects the mode whose electric field is polarized parallel to the fins, while allowing the other mode to propagate through and reflect off the end of the cavity. By orienting the fins orthogonally, one mode is confined between the left end of the cavity and the left edge of the right set of fins, while the other is confined between the right edge of the left set of fins and the right end of the cavity, as shown in Fig.~\ref{fig:modes_2_cylinder}. 

As a result, though both modes continue to see a cavity of equal effective length, the fins offset their profiles by a compensating $\lambda/4$, so that now $\tilde{\v{E}}_1^*$ and $\tilde{\v{B}}_0$ have the same phase within the cavity. Furthermore, since these are overmoded balanced hybrid modes, $\tilde{\v{E}}_1^*$ and $\tilde{\v{B}}_0$ are approximately parallel to each other, with the same spatial distribution, maximizing the signal form factor. Note that within each set of fins, one mode is present (and roughly doubled in magnitude) while the other is exponentially suppressed. Conservatively assuming that these regions contribute zero to the signal form factor, we roughly estimate $C_{\text{sig}} \approx 19/21 \approx 0.9$.

\subsection{Optimized Design}
\label{subsec:optimized}

\begin{figure}[t!]
\centering
\subfloat[\label{fig:square_cross}]{\includegraphics[width=0.6\linewidth]{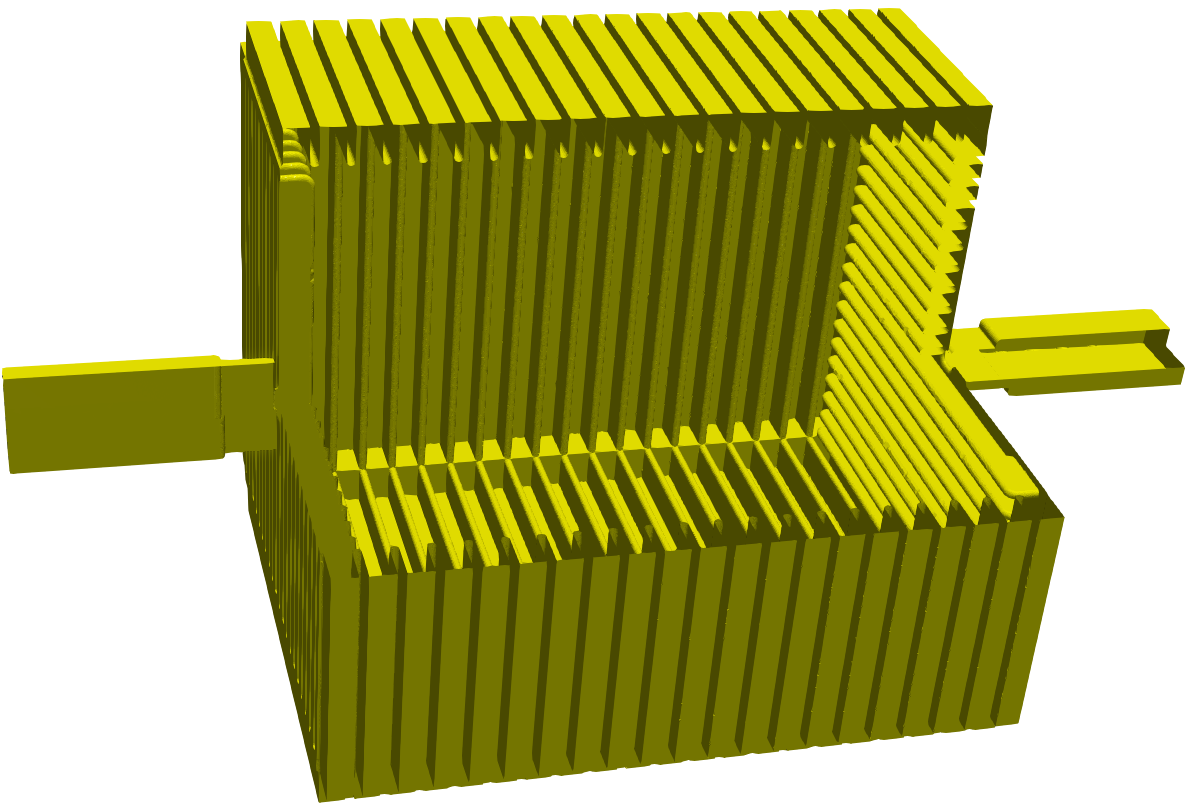}}
\subfloat[\label{fig:eb_square}]{\includegraphics[width=0.38\linewidth]{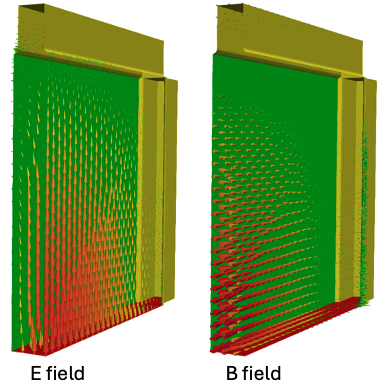}} \newline
\subfloat[\label{fig:modes_2_square}]{\includegraphics[width=0.87\linewidth]{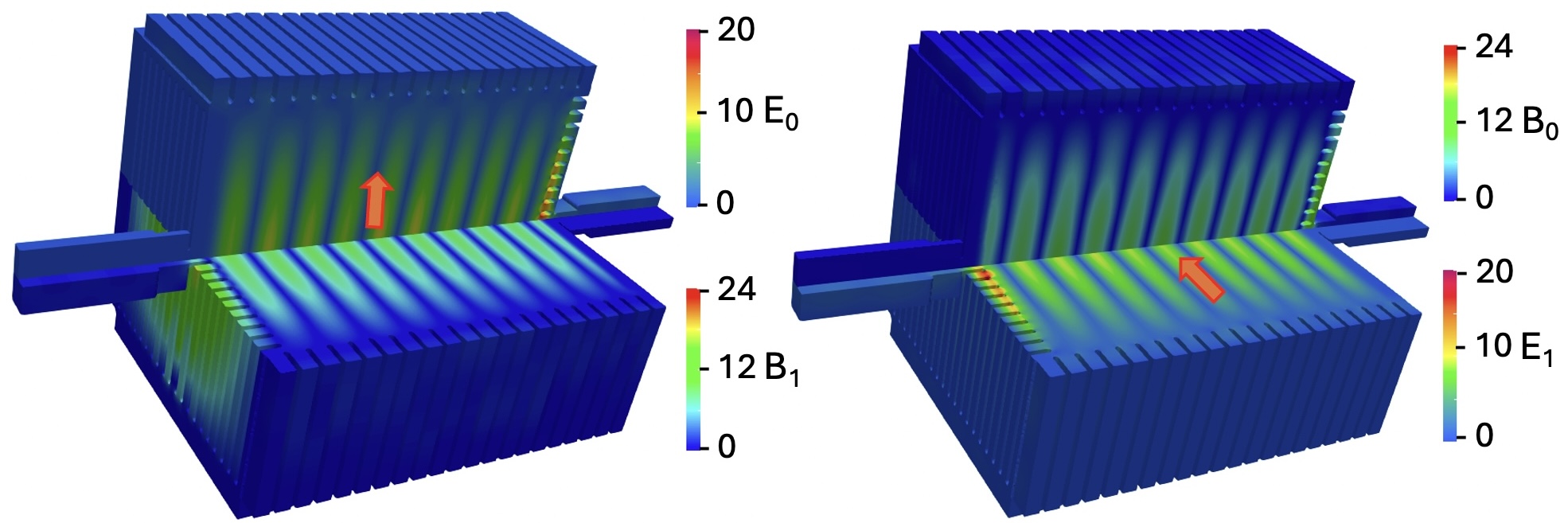}}
\vfill
\caption{The final cavity design, with approximately square cross section. The three subfigures show the same quantities as in Fig.~\ref{fig:cylinder}, and the same useful features of the $\text{HE}_{11}$ modes.}
\label{fig:square}
\end{figure}

\begin{table}
\begin{center}
\begin{tabular}{l|cc|cc}
& & \hspace{-21mm} geometry factor $G^{(s)} \, (\Omega)$ & & \hspace{-18mm} participation ratio $p_s$ \\
surface & tunable & fixed & tunable & fixed \\ \hline
left side plate & $1.40 \times 10^5$ & $7.98 \times 10^4$ & $0.007$ & $0.013$ \\
right side plate & $1.48 \times 10^5$ & $8.04 \times 10^4$ & $0.007$ & $0.013$ \\
bottom side plate & $7.57 \times 10^4$ & $9.41 \times 10^4$ & $0.013$ & $0.011$ \\
top side plate & $7.54 \times 10^4$ & $9.35 \times 10^4$ & $0.013$ & $0.011$ \\
fixed endplate & $5.28 \times 10^3$ & $1.32 \times 10^3$ & $0.186$ & $0.764$ \\
tunable endplate & $1.27 \times 10^3$ & $5.33 \times 10^3$ & $0.774$ & $0.190$ \\ \hline
full cavity & $9.83 \times 10^2$ & $1.01 \times 10^3$ & $1$ & $1$
\end{tabular}
\caption{Contributions of each side of the cavity to the mode loss. Under our normalization $\int_V |\tilde{B}_i|^2 = 1$, the geometry factor for a given surface satisfies $1/G^{(s)}_i = \int_{s} |\tilde{B}_i|^2 \, dS / (\mu_0 \omega)$, where $\omega$ is the mode angular frequency. The participation ratio $p_s \propto 1/G^{(s)}_i$ gives its fractional contribution to the total loss.}
\label{tab:factors}
\end{center}
\end{table}

Further optimization of the design was performed with the guidance of the multi-physics modeling tool Advanced Computational Electromagnetics 3D Parallel (ACE3P)~\cite{Xiao:2024hss}. ACE3P is a high-fidelity simulation suite developed at SLAC for the virtual prototyping of particle accelerator and RF components. It discretizes partial differential equations using curved tetrahedral meshes, enabling accurate modeling of complex geometries. ACE3P contains a wide set of specialized multi-physics solvers; we used Omega3P to calculate the cavity modes and S3P to calculate the $S$-parameters. The solvers were designed for massively parallel computing architectures, and were run on the supercomputers at the National Energy Research Scientific Computing Center (NERSC).

In our final design, shown in Fig.~\ref{fig:square_cross}, the cavity cross-section was changed from circular to approximately square. This was done because corrugated square cavities are easier to fabricate, and they also carry a linearly polarized $\text{HE}_{11}$ mode with the same useful properties~\cite{clarricoats1984corrugated}. We chose the cross-section of the interior of the cavity to be slightly rectangular, with a width of $0.474 \, \mathrm{m}$ and a height of $0.458 \, \mathrm{m}$. This is done to break the degeneracy of the $\text{HE}_{11}$ modes, as well as to separate them from the unwanted $\text{EH}_{11}$ modes, which would also be degenerate for a square cavity. (As a rough estimate, for a side length $d$ and difference $\delta$, the fractional shift in frequency scales as $(\delta/d)^2$, which is $\sim 10^{-3}$ for the parameters chosen.)

The wavelength of the hybrid mode is $\lambda = 105.0 \, \mathrm{mm}$, and the interior of the cavity has length $0.556 \, \mathrm{m}$. There are $21$ corrugations on the side walls, with depth $28.0 \, \mathrm{mm}$ on the top and bottom and $34.4 \, \mathrm{mm}$ on the sides; these are slightly higher than the approximate theoretical value of $\lambda/4$, but were found to result in the best quality factor. The spacing between the side wall corrugations can be varied, and we chose $\lambda/4$, as this ensures a smooth transition with the endplate fins which necessarily have depth $\lambda/4$. Finally, the spacing between the endplate fins is set by a tradeoff between completely reflecting one mode and reducing losses for the other mode. It was set to $20.0 \, \mathrm{mm}$ for the fixed endplate. Since the modes are more highly perturbed near the tunable endplate, we chose a slightly smaller spacing of $19.2 \, \mathrm{mm}$ for it, as this was found to help suppress the mode cross-coupling.

The profiles of the hybrid modes are shown in Figs.~\ref{fig:eb_square} and~\ref{fig:modes_2_square}. Table~\ref{tab:factors} shows how each side of the cavity contributes to the loss; as expected for hybrid modes, losses at the side plates are very small, so that losses at the endplates dominate. For a cavity entirely made of room temperature copper, the ``fixed mode'' has an intrinsic quality factor of $Q = 7.0 \times 10^4$, while the ``tunable mode'' has a slightly lower expected quality factor of $Q = 6.3 \times 10^4$ due to the presence of the aluminum tuning plate, to be discussed below. The fixed mode has a frequency of $2.8565 \, \mathrm{GHz}$, while the frequency of the tunable mode depends on the location of the tuning membrane.

We did not calculate an exact value for the signal form factor, as it involves a volume integral which is nontrivial to perform within ACE3P; however, since the modes are qualitatively similar to the cylindrical case, we estimate $C_{\text{sig}} \simeq 0.9$.

\subsection{Coupling and Frequency Tuning}
\label{subsec:tuners}

\begin{figure}[t!]
\centering
\includegraphics[width=0.7\linewidth]{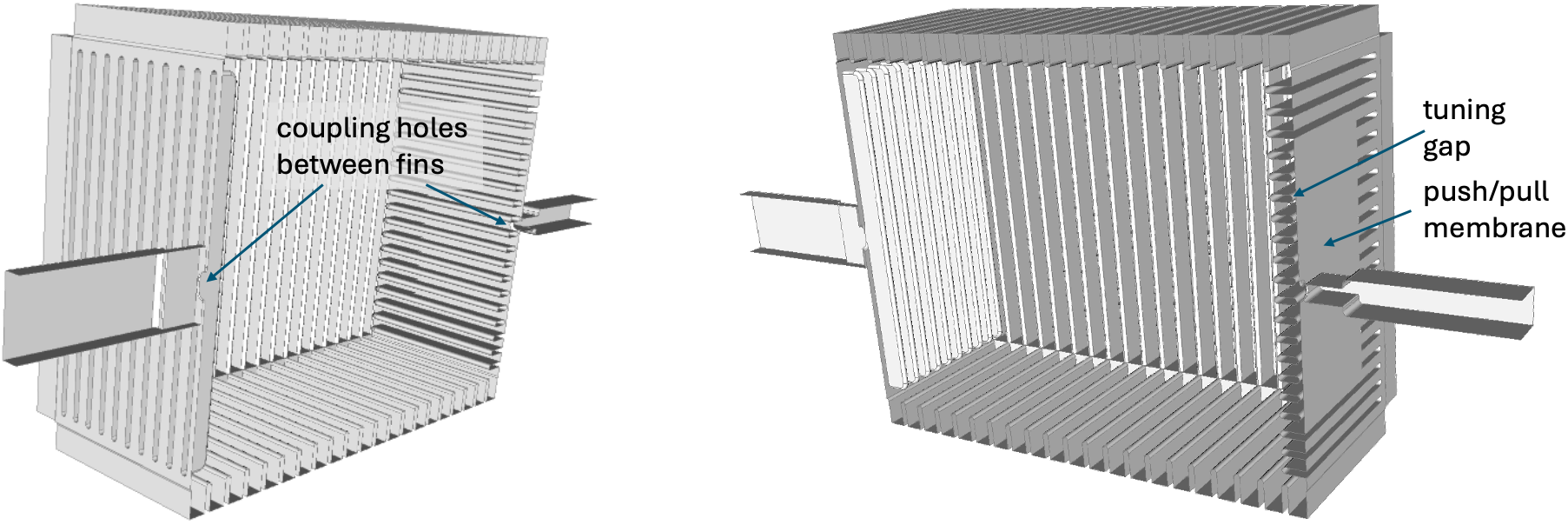}
\vfill
\caption{Schematic depiction of the waveguide couplers and frequency tuning mechanism.}
\label{fig:couplertuner}
\end{figure}

\begin{table}
\begin{center}
\begin{tabular}{ll|cc|cc}
& & & \hspace{-14mm} in bulk & & \hspace{-14mm} on surface \\
mode & coupler & $E_{\mathrm{max}}$ & $B_{\mathrm{max}}$ & $E_{\mathrm{max}}$ & $B_{\mathrm{max}}$ \\ \hline
fixed & large & 9.2 & 9.3 & 25.3 & 28.2 \\
fixed & none & 8.9 & 9.0 & 18.8 & 17.4 \\
tunable & none & 9.3 & 9.3 & 19.2 & 18.0 \\
\end{tabular}
\caption{Maximum magnitudes of the field profiles within the bulk of the cavity and on its surface, in units of $\text{m}^{-3/2}$. The profiles are normalized so that $\int_V |\tilde{E}_i|^2 = \int_V |\tilde{B}_i|^2 = 1$, so that their rms value is $1/\sqrt{V} = 2.9 \, \mathrm{m}^{-3/2}$. The fields are largest on the endplate fins, as shown in red in Fig.~\ref{fig:square}. They are also further enhanced when a large coupling hole is present, which is only required for a non-superconducting setup.}
\label{tab:fields}
\end{center}
\end{table}

RF coupling to the fixed and tunable mode is achieved via magnetically coupled rectangular waveguides on each endplate, which couple to the magnetic fields between the fins. At each endplate, the magnetic field of one mode is large, while the magnetic field of the orthogonally polarized mode is evanescent and therefore exponentially suppressed, automatically suppressing the cross-coupling. Moreover, the waveguide itself is inherently polarization-selective: it couples efficiently only to magnetic fields aligned parallel to its wider dimension, further suppressing the cross-coupling. 

The couplers are designed for critical coupling in the prototype cavity, i.e. RF coupling $\beta_i = 1$, and therefore require an external quality factor of approximately $7.0 \times 10^4$. To minimize mode perturbations and field enhancements of the overmoded $\text{HE}_{11}$ mode due to the coupling hole, a quarter-wave transformer is incorporated into the coupler waveguide. This transformer enhances the coupling efficiency, allowing for a smaller coupling hole with less impact on the mode structure. The coupler design used in the prototype cavity is shown at left in Fig.~\ref{fig:couplertuner}. 

The maximum values of the field profiles are given in Table~\ref{tab:fields}. Within the bulk of the cavity, they are largest on the cavity axis, while on the endplate fins they are roughly enhanced by a factor of $2$. The maximum field on the endplates is further enhanced when large coupling holes are introduced. In a superconducting cavity, the required coupling hole size for critical coupling would be much smaller, removing this enhancement. However, for the simulation results below we will continue to assume large coupling holes.

The tunable mode frequency can be controlled via a tuning gap controlled by a tunable membrane, as shown at right in Fig.~\ref{fig:couplertuner}. This tuning gap is located behind the end-plate fins, making it effective only for the ``tunable'' $\text{HE}_{11}$ mode that can propagate into the fins, while having negligible effect on the orthogonally polarized ``fixed'' mode. Displacing the membrane through a total range of $2 \, \mathrm{mm}$ would change the effective length of the cavity seen by the tunable mode by 0.4\%, which corresponds to a tuning range of $11 \, \mathrm{MHz}$. In addition, we checked that such tuning would preserve the integrity of the $\text{HE}_{11}$ modes.

\subsection{Cross-Coupling Suppression}
\label{subsec:cross_couple}

\begin{figure}[t!]
\centering
\subfloat[\label{fig:rotation}]{\includegraphics[width=0.42\linewidth]{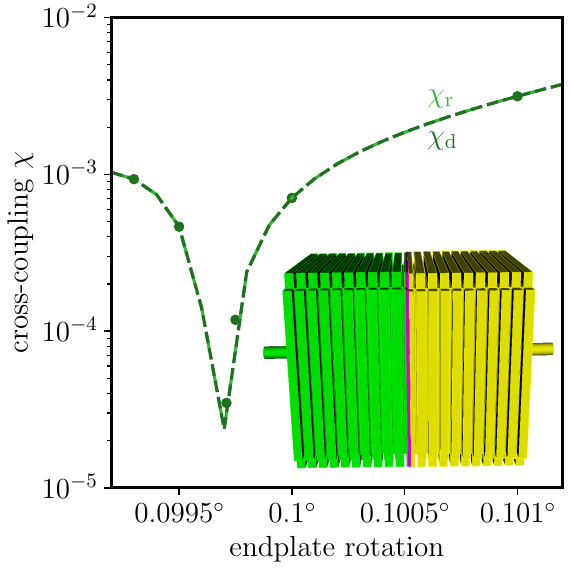}} \quad
\subfloat[\label{fig:skew}]{\includegraphics[width=0.42\linewidth]{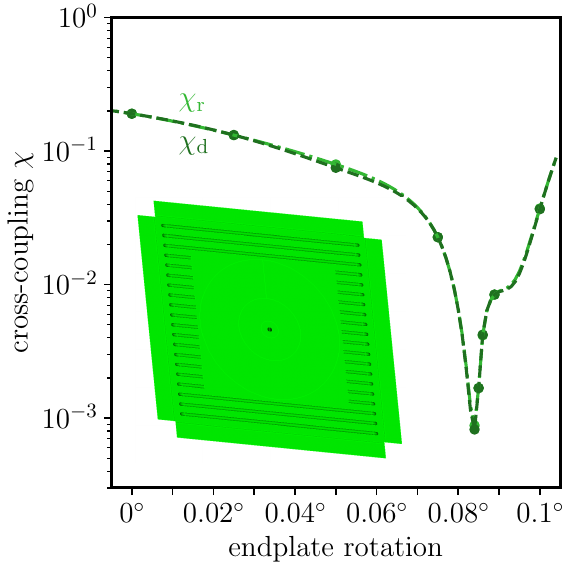}}
\vfill
\caption{Simulations of cross-coupling suppression. Imperfections of the cavity geometry lead to a larger $\chi_{\mathrm{d}}$ and $\chi_{\mathrm{r}}$, which can be simultaneously suppressed by rotating an endplate. Dots show simulated results, and the dashed curve is an interpolation. To demonstrate the point, we consider simulated imperfections of the cavity which are much more severe than what would actually occur. (a) Half of the cavity is twisted by $0.1^\circ$. Rotating the endplate by a similar amount restores a low value of the cross-coupling, $\chi \lesssim 10^{-4}$. (b) The cavity is skewed, so that one diagonal is $0.8 \, \mathrm{mm}$ longer than the other. Rotating the endplate restores $\chi \lesssim 10^{-3}$.}
\label{fig:suppression}
\end{figure}

The cavity with two couplers is a two-port network, with each port dominantly coupled to one of the orthogonal $\mathrm{HE}_{11}$ modes. As described in Eq.~\eqref{eq:cross_coupling_def}, the scattering parameter $S_{\mathrm{rd}}$ is used to quantify the cross-coupling. In the ideal cavity geometry the cross-coupling vanishes exactly, and this remains true when the cavity is tuned in an azimuthally uniform way. However, in practice cross-coupling can be introduced by fabrication imperfections and misalignments. Some of these effects can be mitigated by incorporating tuning mechanisms into the cavity, e.g.~by rotating the tunable endplate about the cavity axis.

To check the feasibility of this concept, we used ACE3P to compute the cross-coupling at the fixed mode frequency, for several choices of the cavity geometry. First, we considered a situation where half of the cavity is twisted along the cavity axis by $0.1^\circ$ relative to the other half. This is a much larger imperfection than one would expect in the fabrication process, and it degrades the mode isolation to $\chi \sim 10^{-1}$. However, as shown in Fig.~\ref{fig:rotation}, rotating the endplate can restore the isolation to $\chi \lesssim 10^{-4}$. 

This result might not be particularly surprising, as both the simulated imperfection and the endplate adjustment involve rotation about the same axis. As a more extreme example, we considered a skew of the entire cavity, so that its diagonals are lengthened and shortened by $0.4 \, \mathrm{mm}$. Again, this is a substantially larger change than one would expect from fabrication, and it degrades the mode isolation to $\chi \sim 10^{-1}$. As shown in Fig.~\ref{fig:skew}, by rotating the endplate, the simulated mode isolation can be recovered to $\chi \sim 10^{-3}$. 

\subsection{Mechanical Mode Mixing Form Factors}
\label{subsec:form_factors}

In a full treatment of mechanical mode mixing, one would compute the form factor $\eta_p$ in Eq.~\eqref{eq:eta_mix} for each mechanical mode profile $\bm{\xi}_p$. However, computing these profiles would be difficult for our complex cavity geometry, and the results would be different for a superconducting version of the cavity. Therefore, in this work we will simply calculate $\eta_p$ for some simple choices of $\bm{\xi}_p$. 

Specifically, we consider deformations where one plate moves inward uniformly. For example, for the bottom plate, we take $\bm{\xi}_p = \hat{\v{z}}$ on that plate and zero elsewhere on the cavity. For the four side plates, we find 
\begin{equation} \label{eq:side_plate_eta}
\eta_p \simeq \begin{cases} 2.8 \times 10^{-4} & \text{left} \\ 3.4 \times 10^{-4} & \text{right} \\ 3.6 \times 10^{-4} & \text{top} \\ 1.8 \times 10^{-4} & \text{bottom}  \end{cases}
\end{equation}
while for the two endplates we find
\begin{equation}
\eta_p \simeq \begin{cases} 7.0 \times 10^{-3} & \text{tunable} \\ 6.8 \times 10^{-4} & \text{fixed} \end{cases}.
\end{equation}
These numbers do not correspond directly to a specific normalized mechanical mode, but they give a rough indication of the typical size of $\eta_p$ for low-lying mechanical modes. 

We can read off several qualitative conclusions from these results. First, $\eta_p$ is highly suppressed for the side walls, which is sensible because the fields of both modes are small there. Note that the values in Eq.~\eqref{eq:side_plate_eta} have an order-one variation, even though the cavity is left/right and top/bottom symmetric, because it is difficult to calculate such small $\eta_p$ precisely. This is both because it depends sensitively on the direction of $d\v{S}$ for the mesh elements in each corrugation, and because the field profile overlaps in Eq.~\eqref{eq:eta_mix} oscillate in sign.

Next, $\eta_p$ is much larger on the endplates, but still much less than one. This occurs because near each endplate, the magnetic and electric fields of one of the modes becomes large, as shown in Fig.~\ref{fig:modes_2_square}. However, their overlap with the magnetic and electric fields of the other mode is still suppressed because of the $\lambda/4$ offset. Note that $\eta_p$ is significantly higher for the tunable endplate, since the tuning mechanism perturbs the mode profiles nearby. This suggests that in a full-scale experiment, it would be important to stiffen the tunable endplate to reduce its vibrations, though this may be in tension with achieving a wide tuning range. 

Since the surface fields are not suppressed at the endplates, it is possible for a mechanical mode to have a significantly higher, order-one form factor, if its oscillations in sign precisely match those of the mode overlap in the integrand. However, we expect that such an effect could only occur in a small part of the frequency range, for a few special modes. Since the mechanical modes in a superconducting version would be different, we defer this question to future study.

\section{Mechanical Design and Fabrication}
\label{sec:mech_fab}

\subsection{Cavity Assembly}

\begin{figure}
\centering
\subfloat[\label{fig:mech_cross}]{\includegraphics[width=0.55\linewidth]{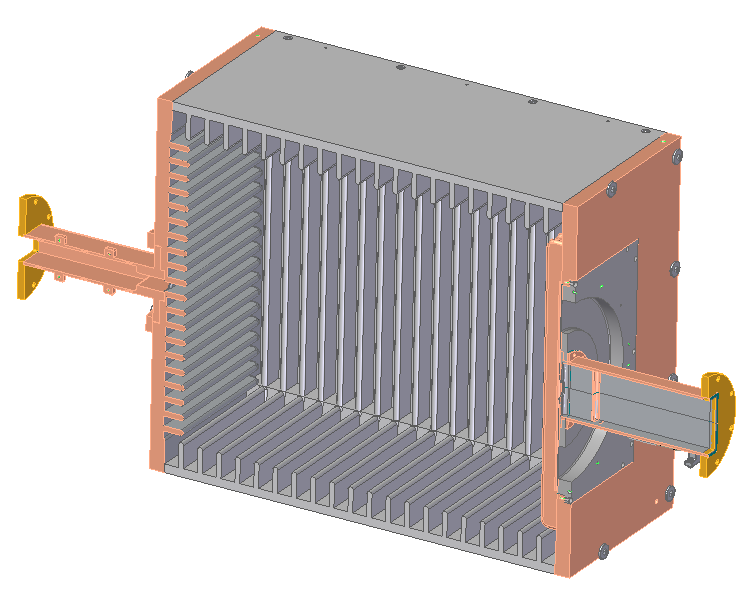}}
\subfloat[\label{fig:mech_full}]{\includegraphics[width=0.45\linewidth]{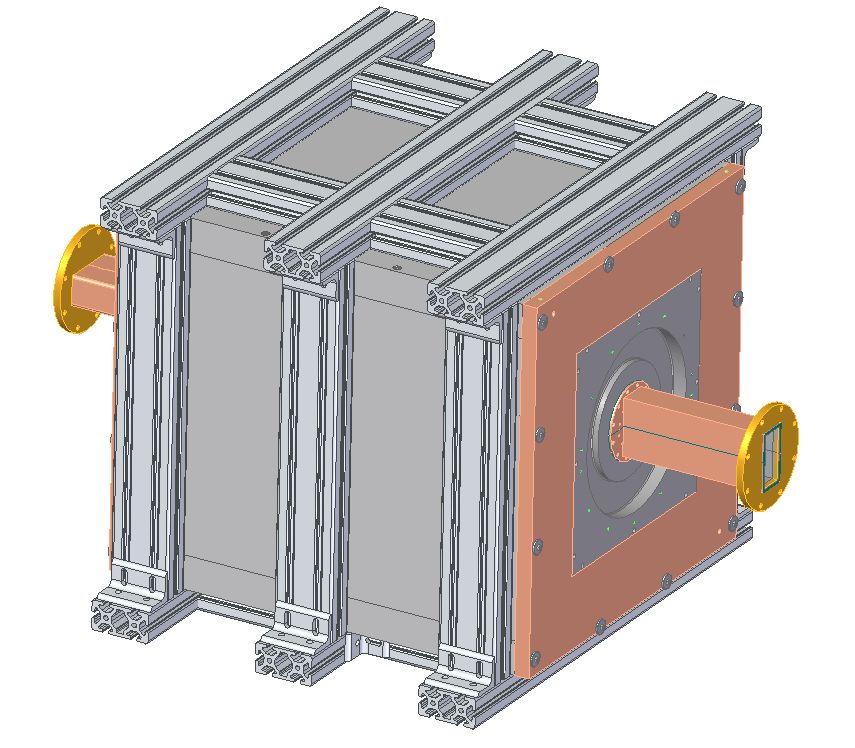}}
\vfill
\caption{Mechanical design of the cavity. (a) Cross section. The cavity is assembled from four corrugated aluminum side walls, two corrugated copper endplates, an aluminum tuning plate on one endplate, and waveguide couplers. (b) Render of the assembled cavity and support structure.}
\label{fig:mech}
\end{figure}

The original cylindrical cavity concept would be difficult to fabricate from niobium. One option would be to cut the cylinder body out from a single piece of niobium, but such an ingot would be impractically large and expensive. Another option would be to make $21$ separate thin rings and stack them together. However, fabrication and handling of these parts to the required tolerance would be very challenging. 

By contrast, the final rectangular cavity design can be assembled from six flat plates, as shown in Fig.~\ref{fig:mech}. Compared to the rings that would be used for a cylindrical cavity, these larger pieces are less fragile, and easier to align. Due to budget constraints, the plates were made of nonsuperconducting materials. To minimize losses, the endplates were made of OFE copper C101 due to the higher fields there. The side plates were made of high conductivity aluminum 6101-T64, and the plate containing the tuning membrane was made of aluminum 6061 due to its good mechanical properties. 

Since the cavity is highly overmoded, the modes are sensitive to deviations of the walls from their ideal geometry, especially on the endplate fins. We demanded that the individual parts were machined flat to at least $\sim 50 \, \mu\mathrm{m}$ precision. Although such tolerance is tight, it is well within the reach of a good CNC center, provided the metal is heat treated before and after machining to remove residual stress. 

Before assembly, the parts went through a metrology verification and local modification. The left and right side plates were mounted on the top and bottom side plates though $1/4$ inch dowel pins with press fit tolerance. After the dowel pins were engaged, the side plates were tightened together using a set of $1/4$ inch socket head screws through clearance holes. This central barrel was supported by a frame made of standard 80/20 aluminum extrusions and mounted on an optical table. 

The fixed endplate is directly mounted on the cavity barrel, but the tunable endplate is mounted against two jack screws with a range of motion of $\pm 250 \, \mu\mathrm{m}$, whose heights can be used to perform a roll angle adjustment, i.e.~to rotate this endplate about the cavity axis. Once the desired orientation is found, the endplate is tightened against the central barrel using wide flange screws. One could also adjust the yaw angle of the tunable endplate by controlling the differential height of additional jack screws, but this was not pursued in the current prototype. 

\subsection{Tuning Mechanism}
\label{subsec:cavity_tuning}

\begin{figure}
\centering
\subfloat[\label{fig:tuner_fab_in}]{\includegraphics[width=0.474\linewidth]{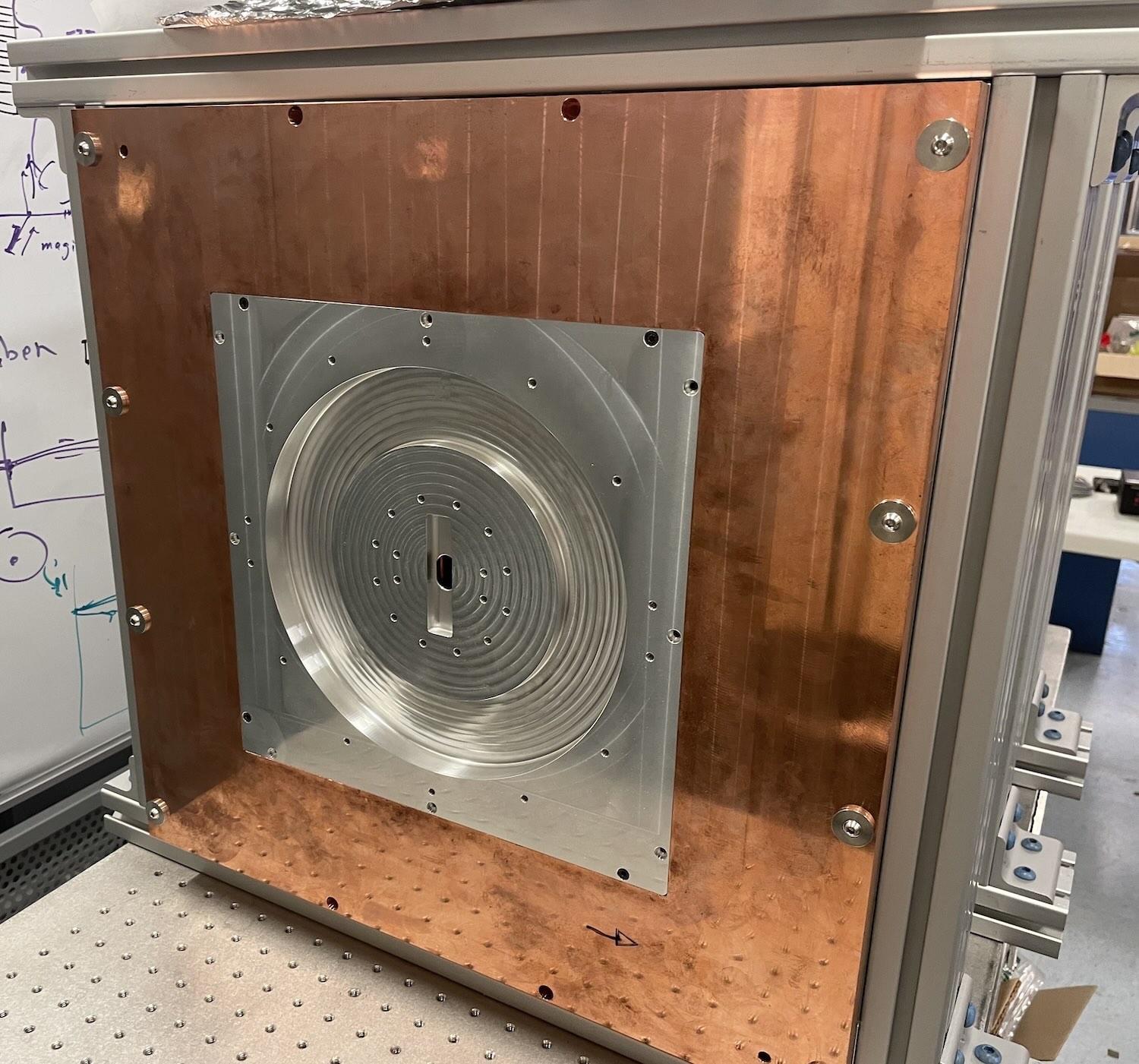}} \quad
\subfloat[\label{fig:tuner_fab_out}]{\includegraphics[width=0.45\linewidth]{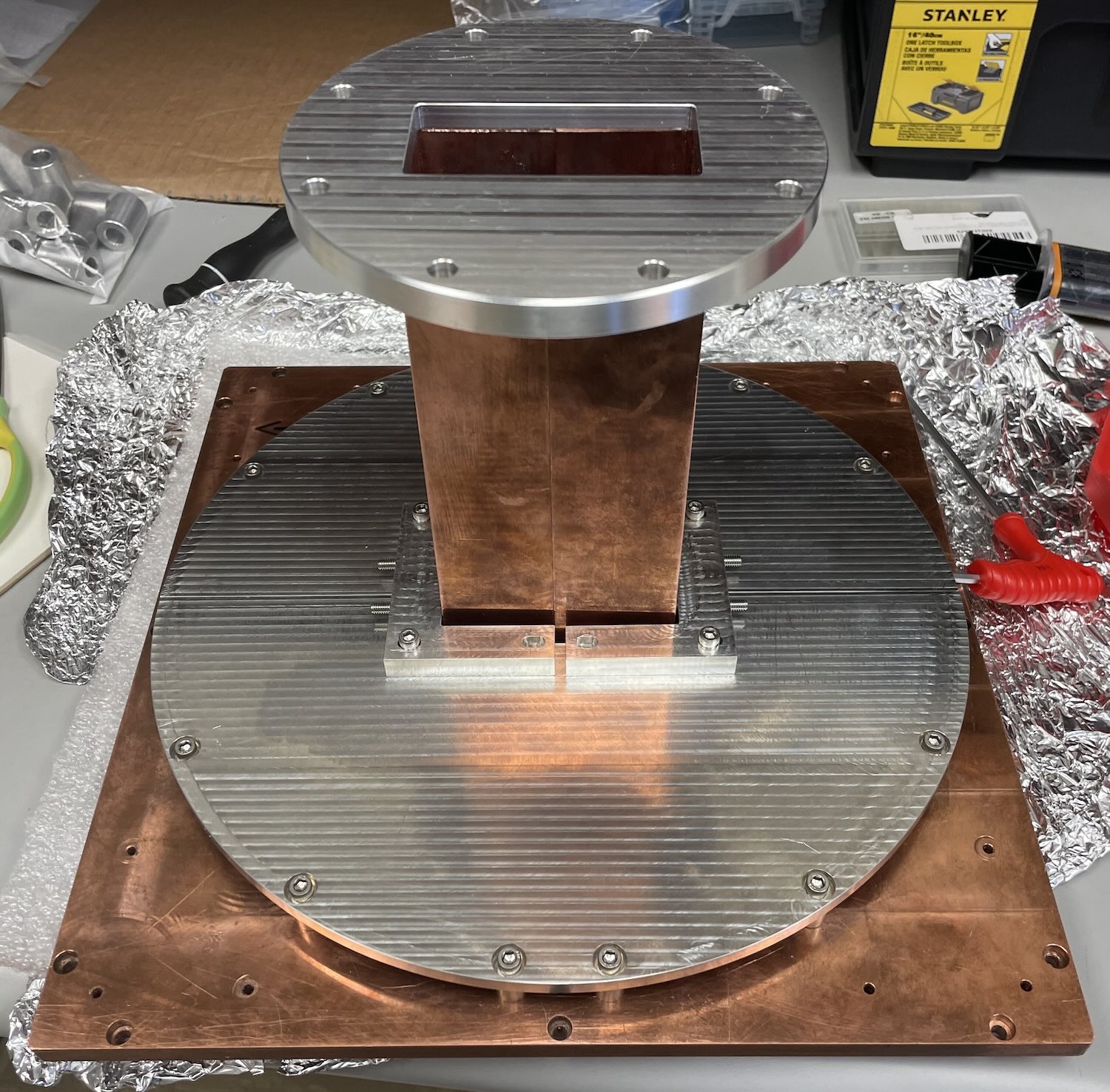}}
\vfill
\caption{The tuning mechanism. (a) The inner part of the tuning plate is mounted on the tunable endplate, and contains a thin circular tuning membrane. (b) The outer part of the tuning plate, with waveguide coupler. Rotating screws near the center causes this part to push on the tuning membrane, deforming it inward.}
\label{fig:tuner_fab}
\end{figure}

\begin{figure}[htb]
\centering
\subfloat[\label{fig:deformation}]{\includegraphics[width=0.35\linewidth]{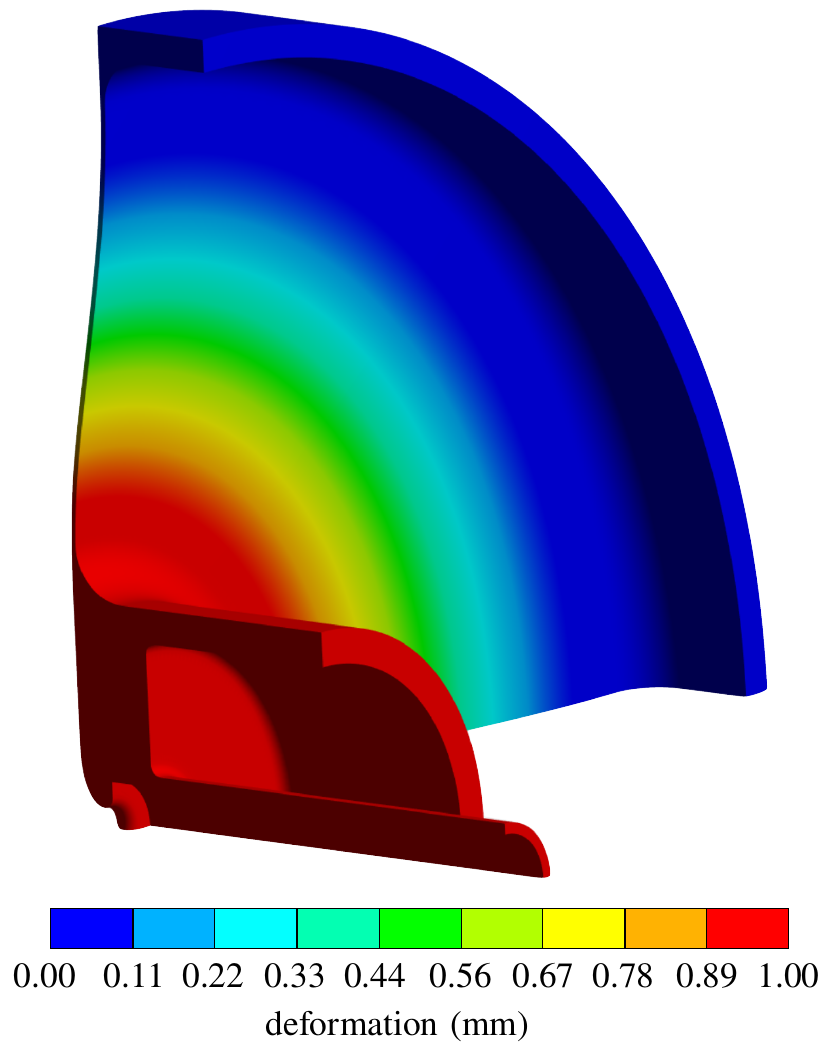}} \quad
\subfloat[\label{fig:stress}]{\includegraphics[width=0.35\linewidth]{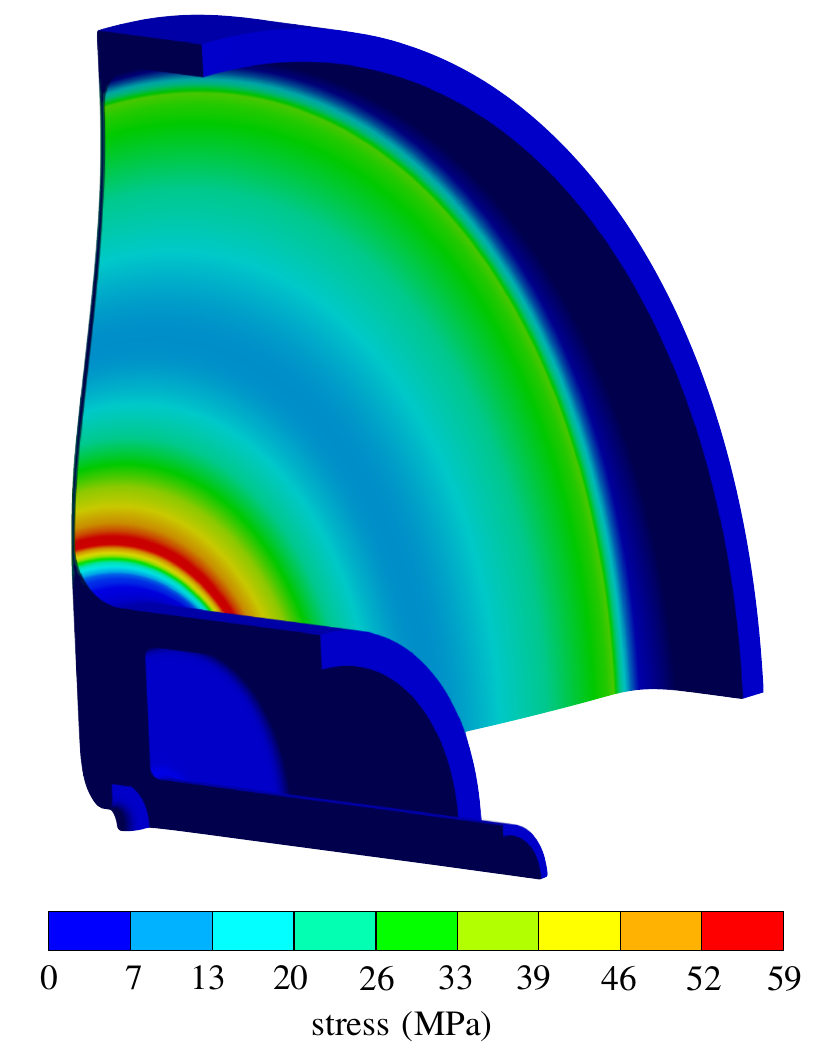}}
\vfill
\caption{Mechanical analysis of the tuning membrane. This figure shows an earlier tuning concept with a central handle that can be pushed or pulled. The final design, shown in Fig.~\ref{fig:tuner_fab_out}, is expected to have similar deformation and stress. (a) Deformation of the membrane when pushed $1 \, \mathrm{mm}$ inward. (b) Equivalent (von Mises) stress of the membrane in this position.}
\label{fig:tuner_sim}
\end{figure}

As shown in Fig.~\ref{fig:cavity_inside}, the tunable endplate contains a large square opening, into which is placed an aluminum tuning plate with a thin circular membrane of $1 \, \mathrm{mm}$ thickness, as shown in Fig.~\ref{fig:tuner_fab_in}. The membrane was designed so that it could be deformed through a total range of $2 \, \mathrm{mm}$, to yield the desired $\sim 10 \, \mathrm{MHz}$ tuning range discussed in Sec.~\ref{subsec:tuners}. The membrane is initially bowed outward, and in its innermost position, it is roughly flush with the wall of the tunable endplate, so that the tunable mode frequency is approximately equal to that of the fixed mode. 

To fabricate the membrane, we considered brazing a circular piece of copper foil of $1 \, \mathrm{mm}$ thickness onto the tuning plate. However, this approach could not achieve adequate planarity, due to the wrinkling of the foil in the heat. Therefore, we machined the tuning membrane directly in the bulk of a thicker aluminum plate. To check the feasibility of this mechanism, we simulated the stress and strain in the tuning membrane. As shown in Fig.~\ref{fig:tuner_sim}, the maximum equivalent (von Mises) stress in the membrane is less than half of the maximum allowed value for aluminum, which is $130 \, \mathrm{MPa}$.

We considered multiple concepts for deforming the membrane. One is shown in Fig.~\ref{fig:tuner_sim}, but the final design, shown in Fig.~\ref{fig:tuner_fab_out}, involves an outer tuning plate made of two half-circular plates, which can be pushed against the tuning membrane by rotating four screws. This design was simple to implement, but has the disadvantage that it can only push the tuning membrane inward, halving the tuning range to about $5 \, \mathrm{MHz}$. 

\section{Characterization of the Cavity}
\label{sec:cavity_test}

\begin{figure}
\centering
\includegraphics[width=0.6\linewidth]{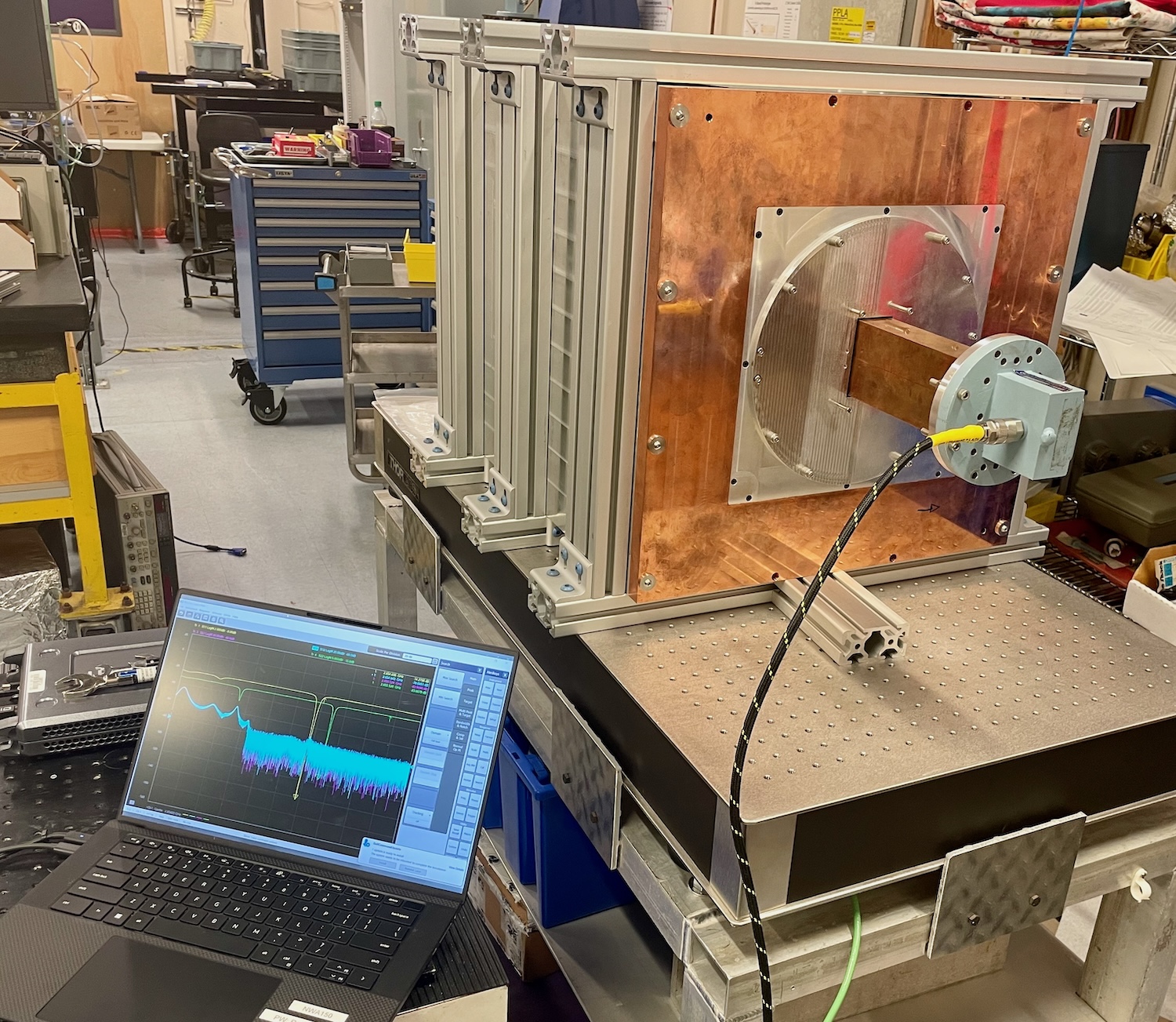}
\vfill
\caption{Setup for bench measurements of the finished cavity.}
\label{fig:bench}
\end{figure}

\begin{figure}
\centering
\subfloat[\label{fig:tuning_measurement}]{\includegraphics[scale=0.82]{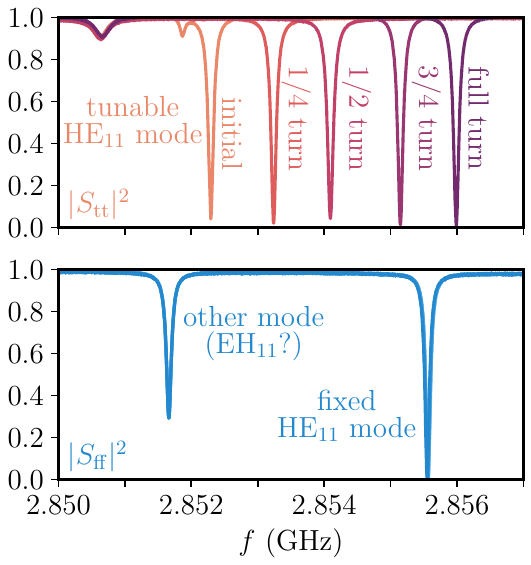}} 
\subfloat[\label{fig:coupling_measurement}]{\includegraphics[scale=0.82]{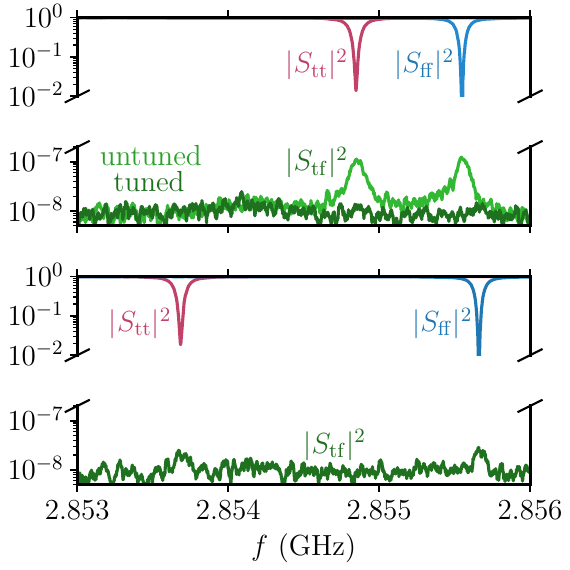}} 
\vfill
\caption{Measurements of the microwave scattering parameters. (a) Demonstration of the tuning range. As the tuning screws were rotated, the frequency of the tunable $\mathrm{HE}_{11}$ mode increased by about $4 \, \mathrm{MHz}$, while the frequency of the fixed $\mathrm{HE}_{11}$ remained the same. (b) Demonstration of decoupling. Top: at a point in the middle of the tuning range, the cross-coupling $|S_{\mathrm{tf}}|^2$ was originally $\sim 10^{-7}$. By rotating the tunable endplate, it was reduced to $\sim 10^{-8}$, the noise floor of the analyzer. Bottom: the decoupling procedure was repeated at another frequency, with similar results.}
\label{fig:measurements}
\end{figure}

Bench measurements were performed by connecting a two-port P9373A Keysight network analyzer to the waveguides, as shown in Fig.~\ref{fig:bench}, and measuring the microwave scattering parameters. We discuss mode frequency tuning in Sec.~\ref{subsec:mode_tuning} and cross-coupling suppression in Sec.~\ref{subsec:decoupling}. Throughout, we label the waveguides on the tunable plate and fixed plate by $t$ and $f$, respectively, to remain agnostic about which mode will be the loaded mode and which will be the signal mode.

\subsection{Mode Properties and Frequency Tuning}
\label{subsec:mode_tuning}

By measuring $S_{\mathrm{ff}}(f)$, shown in blue in Fig.~\ref{fig:tuning_measurement}, we inferred that the fixed mode had frequency $2.8556 \, \text{GHz}$. Using the $Q$-circle method, we found that its intrinsic quality factor was $Q_{\text{int}} = 6.87 \times 10^4$ and the external quality factor was $Q_{\text{ext}} = 7.13 \times 10^4$, so that the waveguide was approximately critically coupled. These results are as expected from our simulations. Note that there is a second dip in $|S_{\mathrm{ff}}|^2$ at lower frequency; this is likely an $\text{EH}_{11}$ mode, split by the rectangularity of the cavity. 

In the initial configuration, the tunable mode had a slightly lower intrinsic quality factor of $5.6 \times 10^4$, as expected, and a frequency of $2.8522 \, \text{GHz}$. Each $1/4$ turn rotation of the tuning screws pushed the tuning plate inward by $0.20 \, \mathrm{mm}$. As expected, this increased the tunable mode frequency by about $1 \, \mathrm{MHz}$ at each step, as shown in Fig.~\ref{fig:tuning_measurement}. We conservatively stopped after one full turn, as the tunable mode frequency became higher than that of the fixed mode. Throughout this procedure, the tunable mode's quality factor, the coupling strengths of the waveguides, and all properties of the fixed mode remained essentially unchanged. 

This measurement demonstrates that the tunable mode frequency can be adjusted across a range of $4 \, \mathrm{MHz}$, and brought close to the fixed mode frequency. 

\subsection{Suppression of Cross-Coupling}
\label{subsec:decoupling}

As discussed above, the cross-coupling can be inferred from $|S_{\mathrm{tf}}(f)|^2$ at the frequency of the fixed or tunable mode, and reduced by rotating the tunable endplate. As shown in the top of Fig.~\ref{fig:coupling_measurement}, we adjusted the tuning plate to a point in the middle of the tuning range and measured $|S_{\mathrm{tf}}(f)|^2 \sim 10^{-7}$ at both mode frequencies. As expected, $|S_{\mathrm{ft}}(f)|^2$ was essentially identical, since the network is reciprocal. 

By manually adjusting the jack screws on the tunable endplate, we reduced this value down to $|S_{\mathrm{tf}}(f)|^2 \sim 10^{-8}$. During the process, the frequencies of the modes also shifted slightly, by about one linewidth. To check the robustness of this result, we tuned to a different frequency and repeated the procedure, as shown at the bottom of Fig.~\ref{fig:coupling_measurement}, finding a similar result. 

This measurement demonstrates that a cross-coupling $\chi \sim 10^{-4}$ can be achieved in our prototype cavity, corresponding to 80 dB of noise reduction. Even further reduction may be possible, but would not be visible in our measurement setup given the noise floor of the network analyzer. 

\section{Discussion}
\label{sec:conclusion}

In Sec.~\ref{subsec:related}, we review the history of heterodyne detection, recent developments, and potential next steps. We discuss what could be achieved with a superconducting version of our prototype in Sec.~\ref{subsec:future}, and conclude in Sec.~\ref{subsec:conclusion}. 

\subsection{Related Work and Future Directions}
\label{subsec:related}

The general concept of heterodyne detection was first proposed for gravitational waves in the 1970s~\cite{Braginskii:1973vm,Pegoraro:1977uv,PEGORARO1978165} and prototyped in the 1980s~\cite{REECE1984341}. In these experiments, the dominant signal of a gravitational wave is to induce mode mixing by vibrating the walls. In the early 2000s, the MAGO collaboration developed and tested a two-sphere cavity optimized for this purpose~\cite{Bernard:2000pz,Bernard:2001kp,Bernard:2002ci,Ballantini:2005am}. Recently, renewed theoretical interest in this concept~\cite{Berlin:2023grv,Lowenberg:2023tfu} led to a joint effort of DESY and Fermilab's Superconducting Quantum Materials and Systems Center (SQMS)~\cite{Fischer:2024nte} which characterized the electromagnetic and mechanical modes of the original MAGO cavity. (Ref.~\cite{Fischer:2024nte} found order-one values of $\eta_p$ for $\sim 100$ low-lying mechanical modes with frequencies up to $\sim 10 \, \mathrm{kHz}$, which is larger than the generic expectation. This appears to be due to the $\sim \mathrm{cm}$ scale shifts in the cavity geometry which occurred during storage, and would not be present in a newly fabricated cavity. In addition, the loaded and signal modes in the MAGO cavity have the same geometric profiles, which enhances mode mixing whenever there is an asymmetry between the spheres.)

The modern wave of interest in heterodyne axion detection began in 2020~\cite{Berlin:2019ahk,Lasenby:2019prg,Berlin:2020vrk}, and for the past several years, several experimental groups have been developing prototypes. The SHADE collaboration at Fermilab's SQMS has characterized noise in a driven 9-cell superconducting cavity~\cite{Berlin:2022hfx}, and developed single cell cavities with a small, tunable splitting between the $\mathrm{TM}_{020}$ and $\mathrm{TE}_{011}$ mode~\cite{Giaccone:2022pke}. The group also performed a sensitive search for dark photon dark matter using an unloaded superconducting cavity with fixed frequency~\cite{Cervantes:2022gtv}. The SHANHE collaboration at Peking University extended the latter result by demonstrating mechanical frequency tuning at cryogenic temperatures~\cite{SHANHE:2023kxz,SHANHE:2024tpr}, and is developing superconducting cavities optimized for heterodyne axion detection. The Australian UPLOAD collaboration performed an early search using the frequency technique\footnote{The frequency technique was discussed in Sec.~\ref{sec:heterodyne}. The sensitivity projections for the frequency technique in Ref.~\cite{PhysRevLett.127.019901} are weaker than projections for the power technique, but it was argued in Ref.~\cite{Tobar:2022rko} that the two methods have equivalent sensitivity if the dominant noise source is the noise temperature of the amplifier. The extent to which this holds will depend on the axion mass and the details of the experimental setup.}~\cite{PhysRevLett.126.081803}, and another search using the power technique which demonstrated a $\sim 10\%$ tuning range~\cite{Thomson:2023moc}. These experiments were performed using non-superconducting cavities, but the group is developing superconducting cavities in collaboration with Jefferson Lab. Finally, the CERN Quantum Technology Initiative, in affiliation with the Physics Beyond Colliders group, has recently begun developing superconducting cavities for heterodyne axion detection with non-mechanical tuning mechanisms~\cite{PBC:2025sny}. 

Test runs with these upcoming prototypes will provide a proof of principle of heterodyne axion detection in the next few years, and are projected to explore territory far beyond existing astrophysical bounds. They will not, however, achieve sensitivity either to axion dark matter produced with an order-one misalignment angle, or, more ambitiously, to QCD axion dark matter at $\sim \mathrm{MHz}$ frequencies.

The sensitivity projection in Ref.~\cite{Berlin:2022hfx} gives an estimate of the experimental parameters required to reach these important milestones. This will require a larger-scale investment, at a scale similar to the Dark Matter New Initiatives (DMNI) program sponsored by DOE High Energy Physics in the US, its proposed successor (AGILE), or similar small-project-scale funding.  In that larger context, key results from the efforts described above include the demonstration of the heterodyne detection scheme in a cryogenic environment and characterization of relevant noise sources.  These important inputs are highly complementary to the results of our work, which focused on optimization of the cavity geometry for excellent signal efficiency, minimal mode overlap, and tunability. One plausible path forward for heterodyne detection is a consolidation of current small-scale prototype efforts into a larger collaboration equipped to carry out an appropriately ambitious experiment to reach high value milestones in sensitivity, informed by the results of the ongoing prototype efforts. Key decisions required to push this effort forward include the tradeoffs of different cavity designs and tuning mechanisms.

\subsection{Outlook for a Superconducting Prototype}
\label{subsec:future}

\begin{figure}[t]
\centering
\includegraphics[width=0.7\linewidth]{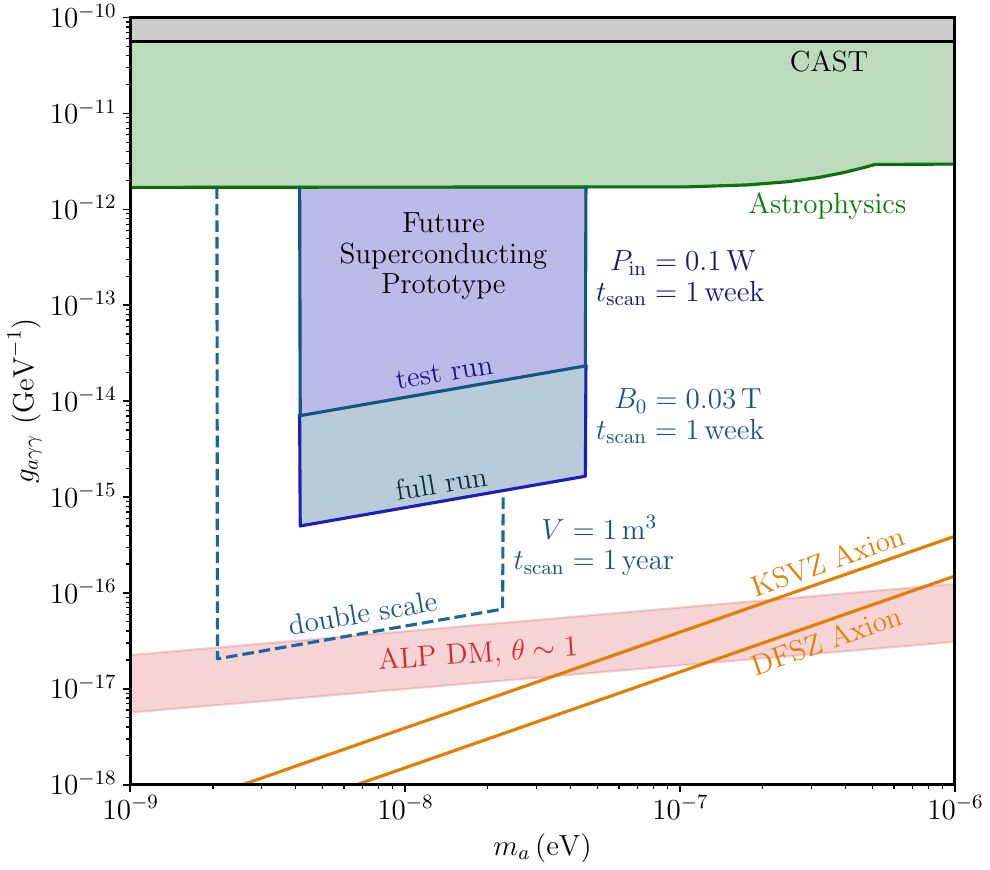}
\vfill
\caption{Our prototype cavity is non-superconducting and operated warm; the resulting low $Q$ and high thermal noise render its sensitivity to axion dark matter negligible. Here we show the projected sensitivity for a hypothetical superconducting cavity with an identical geometry. All parameter values are either already present in the prototype, or would automatically result in a superconducting version made with standard fabrication methods and tested in a standard cryogenic environment; see the text for further discussion. The ``test run'' curve assumes driving with the cavity with a very low input power, while the ``prototype'' curve assumes driving to a moderate magnetic field. The ``double scale'' curve assumes every dimension of the cavity is doubled, with no further changes. We also show existing constraints and theoretical benchmarks: a QCD axion in the KSVZ and DFSZ models, and a (non-QCD) axion produced through standard misalignment with $\theta \sim 1$.}
\label{fig:reach}
\end{figure}

Our prototype has been designed from the ground up to be optimized for heterodyne axion detection. Its high signal form factor $C_{\text{sig}} \simeq 0.9$ compares favorably with the values $C_{\text{sig}} \simeq 0.5$ and $C_{\text{sig}} \simeq 0.2$ achievable in more traditional cylindrical or accelerator cavity designs, respectively, and could yield a substantially higher signal power due to the scaling $P_{\text{sig}} \propto C_{\text{sig}}^2$. Even more radical designs with toroidal cavity geometries have been proposed~\cite{Lasenby:2019prg,Bourhill:2022alm}, though it is less clear how to fabricate and tune them. 

For our design, the natural next step would be to fabricate a superconducting cavity out of solid niobium, niobium-coated copper, or an appropriate mix of the two. The mechanical design, fabrication, and assembly of our prototype have been developed so that many aspects could be carried over directly to superconducting technology, with relevant adaptations of the fabrication details. A similar tuning membrane, cross-coupling suppression mechanism, and waveguide coupler could be used, though one would have to develop motorized mechanisms which can be operated cryogenically. 

Leakage noise measurements could be carried out in an existing cryogenic test facility, and would inform other aspects of the superconducting design. For instance, typical niobium cavities have a thickness of a few mm, while our current design has endplates of $20 \, \mathrm{mm}$ thickness. Thicker walls of niobium would certainly be feasible and would greatly reduce vibrations, but may make thermal management more difficult. Moreover, stiffening the tuning membrane might lower the tuning range. The appropriate tradeoff can be determined by performing \textit{in situ} measurements. 

In an experimental run, one would need to calibrate the axion signal. As with other axion detection concepts, in the heterodyne approach it is not practical to insert a copy of the axion-induced effective current $\mathbf{J}_{\mathrm{eff}}$. However, the signal power is fully determined by the frequencies, quality factors, and external couplings of the cavity modes, which can be extracted from scattering measurements (as discussed in Sec.~\ref{sec:cavity_test}), and the signal form factor $C_{\mathrm{sig}}$ in Eq.~\eqref{eq:signal_form}. This form factor can be found numerically, as we have discussed, or directly measured by mapping out the modes using the ``bead pull'' method, standard in axion experiments. 

To estimate the potential sensitivity, we note that for a thermal noise limited heterodyne axion experiment with amplifier noise at the standard quantum limit, the signal to noise ratio was shown in Ref.~\cite{Berlin:2019ahk} to be
\begin{align} 
\text{SNR} &\sim \frac{\rho_{\mathrm{DM}} \, V}{m_a \, \omega_1} \, (g_{a\gamma\gamma} \, C_{\text{sig}} \, B_0)^2 \sqrt{ \frac{Q_a \, Q_{\text{int}} \, t_e}{T}} \label{eq:optimal_SNR} \\
&= 5 \times \left( \frac{g_{a\gamma\gamma}}{10^{-15} \, \GeV^{-1}} \right)^2 \frac{\rho_{\mathrm{DM}}}{0.4 \, \GeV/\mathrm{cm}^3} \frac{10^{-8} \, \eV}{m_a} \left( \frac{Q_a}{10^6} \right)^{1/2} \nonumber \\
&\qquad \times \frac{V}{0.12 \, \mathrm{m}^3} \frac{2 \pi \, (2.85 \, \mathrm{GHz})}{\omega_1} \left(  \frac{C_{\text{sig}}}{0.9} \right)^2 \nonumber \\
&\qquad \times \left( \frac{B_0}{0.03 \, \mathrm{T}} \right)^2 \left( \frac{Q_{\text{int}}}{3.9 \times 10^{10}} \frac{t_e}{1 \, \mathrm{week}} \frac{2 \, \mathrm{K}}{T} \right)^{1/2}.
\end{align}
This estimate of the SNR holds up to an order-one factor which depends on details such as the precise shape of the axion frequency spectrum, and assumes the frequency scan is performed uniformly in $\log(m_a)$. In the three lines of the second equality, we plug in standard parameters for dark matter at a representative mass and coupling, parameters which are already realized in our prototype cavity design, and parameters which would be reasonable in a future superconducting version of the cavity, as will be discussed in further detail below. Note that Eq.~\eqref{eq:optimal_SNR} implicitly assumes that the readout waveguide is optimally overcoupled by a factor of $T / \omega_1 \simeq 15$, which optimizes the scan rate. 

For the cavity, we assume cooling to $T = 2 \, \mathrm{K}$, which is standard for a helium cryostat. In our prototype design, our simulations indicated that a tuning range of $10 \, \mathrm{MHz}$ is both mechanically feasible and preserves the cavity mode properties. Only about half this tuning range was realized in the prototype cavity, since we could only push the tuning membrane inward. We assume that the motorized tuning mechanism for the superconducting cavity can both push and pull, realizing the full potential tuning range. It is simplest to estimate the sensitivity at high frequencies, where leakage noise is automatically suppressed, so we suppose the mode frequency difference is scanned from $1 \, \mathrm{MHz}$ to $11 \, \mathrm{MHz}$.

As for the intrinsic quality factor $Q_{\text{int}}$, we have seen that a pure copper prototype cavity would have $Q_{\text{int}} = 7 \times 10^4$ and a surface resistance $R_s = \sqrt{\omega_0 / 2 \sigma} = 1.4 \times 10^{-2} \, \Omega$, where $\sigma$ is the standard conductivity of room temperature copper. The surface resistance of niobium depends on frequency, temperature, and the manufacturing process. It can be decomposed as $R_s = R_{\text{BCS}}(f, T) + R_{\text{res}}(f)$, where the BCS contribution is proportional to $f^2$ and the residual resistance is due to impurities, defects, and trapped magnetic flux~\cite{Gurevich_2017}. For the LCLS-II project, commercial vendors use nitrogen doping to reliably produce superconducting cavities with $R_{\text{BCS}}(1.3 \, \mathrm{GHz}, 2 \, \mathrm{K}) = 4.5 \, \mathrm{n}\Omega$ and $R_{\text{res}} = 2.5 \, \mathrm{n}\Omega$~\cite{GONNELLA2018143}. We assume the residual resistance is dominated by trapped flux, so that $R_{\text{res}} \propto f^{1/2}$~\cite{Gurevich_2017,7893741}. This yields $R_s = 25 \, \mathrm{n}\Omega$ at the prototype cavity frequency, which gives a projected quality factor $Q_{\text{int}} = G/R_s = 3.9 \times 10^{10}$. 

Though surface resistances in this range are routinely achieved on production-grade SRF cavities, it may seem challenging to achieve these values on complex surfaces, e.g.~with grooves machined out of a thick niobium sheet. However, similar production techniques have already been demonstrated for HL-LHC crab cavities, with very good results~\cite{ristori_development_2023}. We therefore believe there is no fundamental obstacle for obtaining state-of-the art results on machined surfaces, provided proper surface treatments like electropolishing and possibly nitrogen doping or infusion~\cite{dhakal_nitrogen_2020} are adopted, possibly with some development work.

In a superconducting implementation, the rectangular features in our cavity design could lead to multipacting, which could be alleviated by rounding the corners of the fins. Since addressing this question requires dedicated simulations, we defer a detailed study to future work. 

In Fig.~\ref{fig:reach}, we show the projected reach of a superconducting version of our prototype at $\text{SNR} = 2$. The ``test run'' assumes a total scan time of 1 week and an input power $P_{\mathrm{in}} = 0.1 \, \mathrm{W}$, which is within the reach of a standard low-noise oscillator. In this regime leakage noise would almost certainly be negligible, given the low values of $\chi$ and $\eta$ in our design, the small input power, and the $\gtrsim \mathrm{MHz}$ axion frequency. Note that, as discussed above, the signal mode should be highly overcoupled to the readout waveguide. In the optimal case this would reduce its quality factor to $Q_1 \simeq 2.5 \times 10^9$, which both decreases the time required for the signal to ring up, and simplifies the stabilization of the mode frequencies. 

The ``full run'' curve assumes the same total scan time, but a loaded mode with typical magnetic field $B_0 = 0.03 \, \mathrm{T}$. Maintaining this field requires an input power $P_{\mathrm{in}} = \omega_0 B_0^2 V / 2 Q_{\mathrm{int}} = 20 \, \mathrm{W}$, which is well within the capacity of a vertical test stand. At this input power, Table~\ref{tab:fields} implies a maximum magnetic field of $0.2 \, \mathrm{T}$ and electric field of $60 \, \mathrm{MV}/\mathrm{m}$ at the cavity walls, due to the field enhancement at the endplate fins. At these fields, proper surface treatment will be required to control field emission. 

Finally, the ``double scale'' curve assumes a scan time of 1 year and a direct scale-up of the cavity while maintaining $B_0 = 0.03 \, \mathrm{T}$. This automatically halves the mode frequencies, multiplies the volume by $8$, and increases the quality factor to $2.4 \times 10^{11}$. 

We find that even a test run can probe well beyond astrophysical bounds, with a prototype run probing over an order of magnitude further. The double scale setup can probe the theoretical benchmark of axion dark matter produced with a generic, order-one misalignment angle, though operating it may require the construction of a custom cryomodule. In addition, in any heterodyne setup, it would be straightforward to probe axions of frequency orders of magnitude below MHz simply by tuning the modes closer together. We do not show this region because at kHz frequencies the sensitivity is determined by other noise sources whose magnitude is currently uncertain, e.g.~due to the unknown spectrum of mechanical vibrations. Such uncertainties also affect all other axion experiments operating in this frequency range.

Our projections are conservative, in the sense that they assume no improvements in cavity geometry, fabrication, or readout, but reasonable further improvements would enable sensitivity to the QCD axion. For example, at $m_a = 10^{-7} \, \eV$, the ``double scale'' setup is sensitive to couplings $g_{a\gamma\gamma} \simeq 3.6 \, g_{a\gamma\gamma}^{\text{KSVZ}}$. The gap could be closed by improving the values of $B_0$ and $Q$. For instance, the endplate fins roughly double the magnetic field near them; an improved design may allow a higher magnetic field $B_0 = 0.06 \, \mathrm{T}$. Vacuum heat treatment has already been used to lower the surface resistance of SRF cavities to below $1 \, \mathrm{n}\Omega$~\cite{PhysRevApplied.13.014024}, which could further improve the quality factor by a factor of $10$. These two improvements, or simply an additional scaleup in volume, would enable KSVZ axion sensitivity. These statements are consistent with the early projections made in Ref.~\cite{Berlin:2019ahk}, though that work took a higher $B_0$ and lower $Q$ as baseline. 

\subsection{Conclusion}
\label{subsec:conclusion}

Superconducting cavities are a mature technology with the potential to transform the search for light axion dark matter by enabling heterodyne detection. The vast majority of other axion search proposals use static background fields, and face common R\&D needs to detect the very low axion signal power~\cite{Adams:2022pbo}. For instance, they may require new quantum measurement techniques to probe beyond the standard quantum limit, very large magnets and dilution refrigerators, and new magnet designs to achieve unprecedentedly high background fields. By contrast, the heterodyne approach to axion detection faces a distinct set of challenges. Its potential sensitivity is a result of the inherently higher available signal power, and the steady refinement of superconducting cavities for accelerators over the past decades. 

The development of heterodyne axion detection is in its early stages, and results from Fermilab, Peking University, and CERN will soon give significantly more information about its feasibility. Here, we have demonstrated that a hybrid mode cavity can simultaneously achieve a high signal form factor, low noise form factors, and a wide tuning range, albeit at the cost of a lower maximum magnetic field $B_0$. However, many questions about the optimal operation of a heterodyne experiment remain open, and the ultimate cavity design may be completely different from any previously known. Looking forward, we invite collaboration from experimental groups around the world to resolve these questions. 

\section*{Acknowledgement}

We thank Bianca Giaccone, Tom Krokotsch, Krisztian Peters, and Marc Wenskat for discussions regarding mode mixing in the MAGO cavity. This work was supported by the LDRD grant ``An SRF Cavity for Dark Matter Axion Detection'' at SLAC. RF simulations were performed using the supercomputer resources at the National Energy Research Scientific Computing Center (NERSC), which is supported by the Office of Science of the U.S. Department of Energy under contract DE-AC02-05CH11231. KZ was supported by the Office of Science of the U.S. Department of Energy under contract DE-AC02-05CH11231. The work of RTD was partially supported by ANR grant ANR-23-CE31-0024 EUHiggs. The work of SARE was supported by SNF Ambizione grant PZ00P2\_193322, \textit{New frontiers from sub-eV to super-TeV}.

\bibliographystyle{elsarticle-num}
\bibliography{refs}

\end{document}